\documentclass[preprint,prd,aps,nofootinbib]{revtex4}
\usepackage{graphicx}
\usepackage{diagbox}
\usepackage{color}
\usepackage{slashed}
\makeatletter

\newcommand{\Rmnum}[1]{\expandafter\@slowromancap\romannumeral #1@}
\makeatother
\begin{document}
\title{The electromagnetic decays of $X(3823)$ as the $\psi_2(1^{3}D_{2})$ state and its radial excited states}
\author{Wei Li$^{1,2}$\footnote{watliwei@163.com,corresponding author},
 Su-Yan Pei$^{1,2}$, Tianhong Wang$^{3}$\footnote{thwang@hit.edu.cn,corresponding author},
 Ying-Long Wang$^{1,4}$, Tai-Fu Feng$^{1,2}$
 Guo-Li Wang$^{1,2}$\footnote{wgl@hbu.edu.cn,corresponding author}}

\affiliation{&&${^1}$ Department of Physics and Technology, Hebei University, Baoding 071002, China
\nonumber\\$^{2}$ Key Laboratory of High-precision Computation and Application
of Quantum Field Theory of Hebei Province, Baoding, China
\nonumber\\$^{3}$School of Physics, Harbin Institute of Technology, Harbin 150001, China
\nonumber\\$^{4}$Department of Preschool Teacheers College, Baoding 072750, China}
\begin{abstract}
We study the electromagnetic (EM) decays of $X(3823)$ as the $\psi_2(1^{3}D_{2})$ state by using the relativistic Bethe-Salpeter method. Our results are $\Gamma[X(3823)\rightarrow\chi_{_{c0}}\gamma]=1.2$ keV, $\Gamma[X(3823)\rightarrow\chi_{_{c1}}\gamma]=265$ keV, $\Gamma[X(3823)\rightarrow\chi_{_{c2}}\gamma]=57$ keV and $\Gamma[X(3823)\rightarrow\eta_{_c}\gamma]=1.3$ keV. The ratio ${\cal B}[X(3823)\rightarrow\chi_{_{c2}}\gamma]/{\cal B}[X(3823)\rightarrow\chi_{_{c1}}\gamma]=0.22$, agrees with the experimental data. Similarly, the EM decay widths of $\psi_{_2}(n^{3}D_{_2})$, $n=2,3$, are predicted, and we find the dominant decays channels are $\psi_{_2}(n^{3}D_{_2})\rightarrow\chi_{_{c1}}(nP)\gamma$, where $n=1,2,3$. The wave function include different partial waves, which means the relativistic effects are considered. We also study the contributions of different partial waves.
\end{abstract}
%\pacs{12.60.Jv, 14.80.Cp}
%\keywords{X(3823), Electromagenatic Decay, Bethe-Salpeter Method}
\maketitle
\section{Introduction}
The bound state of charm and anti-charm quarks (charmonium) is significant in our knowledge of quantum chromodynamics (QCD). It is a double-heavy meson, but not heavy enough that its relativistic corrections are still large \cite{G.L.Wang.T.F.Feng.X.G.Wu2020}. Then the charmonium is crucial to test the validity of phenomenological models, such as the quark potential model, which already foresee a rich and meaningful quarkonium spectra \cite{S.G1985}.
More charmonia and charmoniumlike states have been discovered experimentally in the last decade, such as the $X(3872)$ \cite{S.K.Choi2003}, $X(3915)$ \cite{S.U.2010}, $\chi_{_{c2}}(3930)$ \cite{S.Uehara2006}, $\psi(4160)$ \cite{P.Pakhlov2008}, $Y(4260)$ \cite{B.Aubert2005}, $Z_{_c}(3900)$ \cite{M.Ablikim2013} and $Z_{_{cs}}(3985)$ \cite{M.Ablikim2011}, and these new states have stimulated great interests of studies, more details can be found in the review papers \cite{N.B2011,H.X.Chen2016,Y.R.Liu2019,N.B2020}.

Recently, a new bound state $X(3823)$ has been observed, which is considered to be a good candidate for spin triplet $D$ wave charmonium $\psi_{_2}(1^{3}D_{_2})$. The Belle Collaboration first observed $X(3823)$ in the $B\rightarrow \chi_{_{c1}}\gamma K$ decay with a statistical significance of $3.8 \sigma$ \cite{V. Bhardwaj2013}. The BESIII Collaboration confirmed this particle in the process $e^{+}e^{-}\rightarrow \pi^{+}\pi^{-}\chi_{_{c1}}\gamma$ with a statistical significance of $6.2 \sigma$ \cite{M. Ablikim2015} and in process $e^{+}e^{-}\rightarrow \pi^{+}\pi^{-}\psi_2(3823)$ followed by $\psi_2(3823)\to \chi_{_{c1}}\gamma$ with a statistical significance greater than $5 \sigma$ \cite{bes}. Its decay to $\pi^{+}\pi^{-}J/\psi$  also observed by the LHCb Collaboration \cite{R.Aaij2020}. The mass of this particle is measured to be $3821.7 \pm 1.3 \pm 0.7$ MeV, and the decay width is less than $16$ MeV at the $90\%$ confidence level \cite{M. Ablikim2015}.

At present, the experimental data of $X(3823)$ is still relatively sparse. However, the experimental results obtained have raised some theoretical concerns about the properties of the particle. This particle has different production channels, for example, it can be produced in the $B$ meson decay \cite{S.J.Sang2015}, $B_{_c}$ decay \cite{Qiang Li2016}, $Z^0$ decay \cite{C.F.Qiao1997}, also in the $e^+e^-$ annihilation \cite{M.B.V2015}, etc.  For its decays, the $D\bar{D}$ channel is forbidden since its mass is below the $D\bar{D}^{*}$ threshold, hence there is no Okubo-Zweig-Iizuka (OZI)-allowed channel. Therefore, the process of single photon radiation \cite{E.J.E2004}, decay into light hadrons \cite{Z.G.H.2010,T.H.Wang2016} are important. Different models \cite{E.J.E2002,D.E2003,T.B2005,B.Q.Li2009,B.W.H2016,W.J.Deng 2015} have studied the radiative decays of $\psi_{_2}(1^{3}D_{_2})$. These studies show that as the strong candidate of $\psi_{_2}(1^{3}D_{_2})$, instead of the strong decays to light hadrons and the channel $\pi^{+}\pi^{-}J/\Psi$, its dominate decay channel is the radiative decay to $\chi_{_{c1}}\gamma$,
which is partly confirmed by the measured branching-fraction ratio ${\cal B}[\psi_{_2}(1^{3}D_{_2})\rightarrow\chi_{_{c2}}\gamma]/{\cal B}[\psi_{_2}(1^{3}D_{_2})\rightarrow\chi_{_{c1}}\gamma]=0.28^{+0.14}_{-0.11}\pm0.02$ \cite{M. Ablikim2021}. So the radiative transitions are crucial to study the property of $X(3823)$.

Most existing theoretical predictions of the $X(3823)$ EM decay are provided by non-relativistic methods. However, we have found the relativistic corrections are large for charmonia, especially for the higher excited states \cite{gengzk,G.L.Wang.T.F.Feng.X.G.Wu2020}, so it is necessary to study the properties of $X(3823)$ with different methods especially relativistic one. The Bethe-Salpeter (BS) equation is a relativistic dynamic equation used to describe bound state \cite{E.E.S AND H.A.B.1951}. Salpeter equation \cite{E.E.S1952} is its instantaneous version which is suitable for the heavy meson, especially the double-heavy meson. We have solved the complete Salpeter equations for different states, see Refs. \cite{K.C.S AND G.L.Wang2004, G.L.Wang2007} as examples, and we have improved this method to calculate the transition amplitude \cite{C.H.C J.K.Chen G.L.Wang2006} with relativistic wave function as input, where the transition formula is also relativistic. Using this improved BS method, we can get relatively accurate theoretical results, which agree well with the experimental data \cite{G.L.Wang20072,3872,bc2s}.

So in this paper, the $X(3823)$ as $\psi_{_2}(1^{3}D_{_2})$ state is studied by the improved BS method, we will focus on the EM decay processes of $X(3823)$. Besides the dominant channels $\psi_{_2}(1^{3}D_{_2})\rightarrow\chi_{_{c1}}\gamma$ and $\psi_{_2}(1^{3}D_{_2})\rightarrow\chi_{_{c2}}\gamma$, the radiative decays $\chi_{_{c0}}\gamma$ and $\eta_{_c}\gamma$, whose studies are lacking in the literature, are also calculated. We also provide the results of $\psi_{_2}(n^{3}D_{_2})\rightarrow\chi_{_{cJ}}(mP)\gamma$, $\psi_{_2}(n^{3}D_{_2})\rightarrow\eta_{_c}(mS)\gamma$ and $\psi_{_2}(n^{3}D_{_2})\rightarrow\chi_{_{c2}}(mF)\gamma$, with $n=2,3$ and $m=1,2,3$. Where $\chi_{_{c2}}(mF)$ is the $F$ wave dominant $2^{++}$ state, mixed with sizable $P$ and $D$ partial waves \cite{wang2022}.

This paper is organized as follows. In Sec \Rmnum{2}, we show theoretical method to calculate the transition matrix amplitude and the form factors as well as the relativistic wave functions of initial and final states. In Sec \Rmnum{3}, we give the results and compare them with other theoretical predictions and experimental data. Finally, we give the discussion and conclusion.

\section{ THE THEORETICAL CALCULATIONS }
In order to avoid tediousness, we will not introduce the BS equation and Salpeter equation, interested reader can find them in Refs. \cite{E.E.S AND H.A.B.1951,E.E.S1952} or our previous paper, for example, \cite{K.C.S AND G.L.Wang2004}.
\subsection{Transition Amplitude }

Take the EM decay $X(3823)\rightarrow\chi_{_{cJ}}\gamma$ as an example, we show how to use our method to calculate the transition amplitude, which can be written as
\begin{eqnarray}
&&\langle \chi_{_{cJ}}(P_{_f},\epsilon_{_2})\gamma(k,\epsilon_{_0})|X(P,\epsilon_{_1})\rangle=(2\pi)^{4}\delta^{4}(P-P_{_f}-k)\epsilon_{_{0{\xi}}}{\cal M}^{\xi},
\end{eqnarray}
where $\epsilon_{_0}$, $\epsilon_{_1}$ and $\epsilon_{_2}$ are the polarization vectors (tensor) of the photon, initial and final mesons, respectively. $P$, $P_{_f}$ and $k$ are the momenta of initial meson, final meson and photon, respectively.
\begin{figure}[!htb]
\begin{center}
\includegraphics[width=0.47\textwidth]{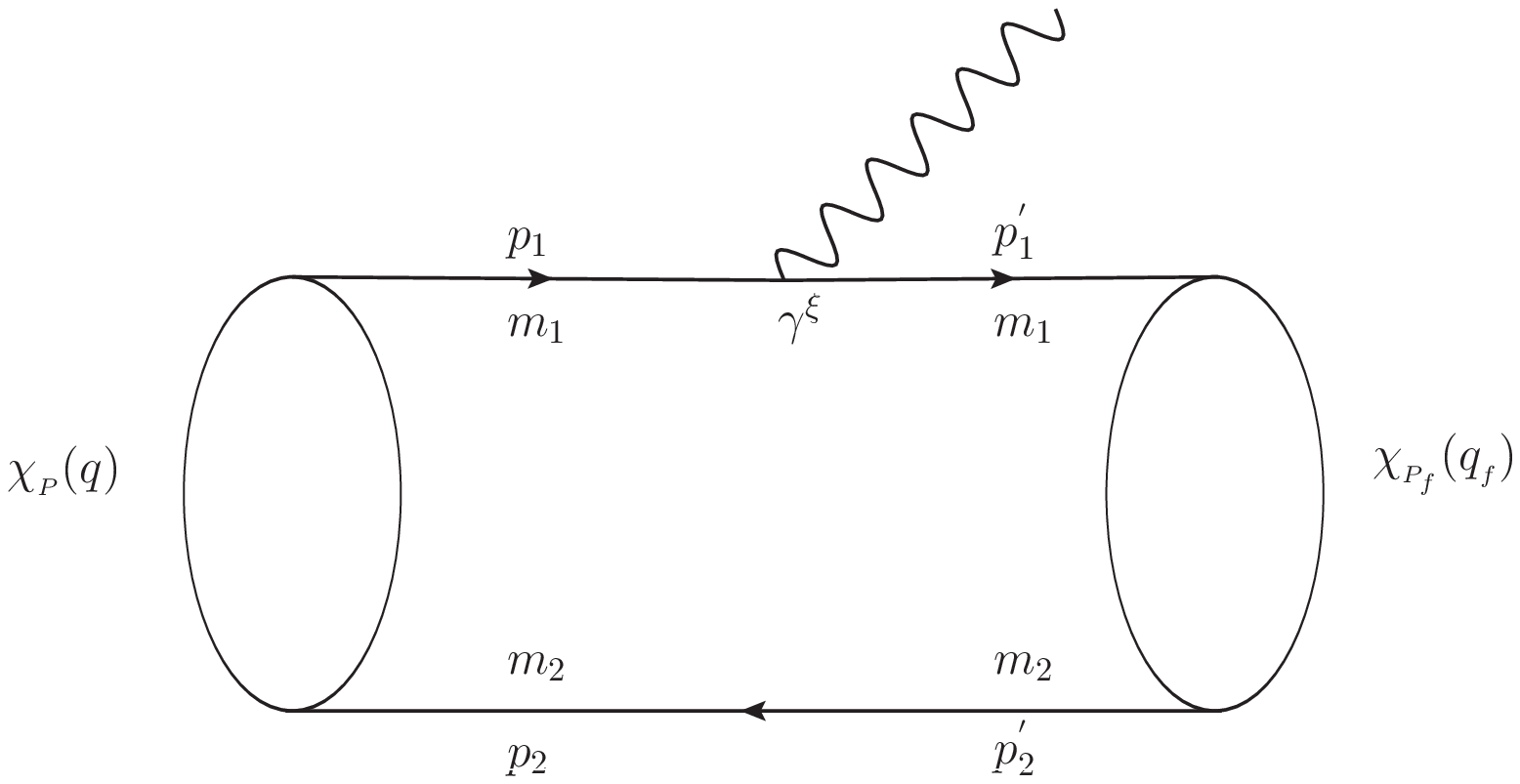}
\includegraphics[width=0.47\textwidth]{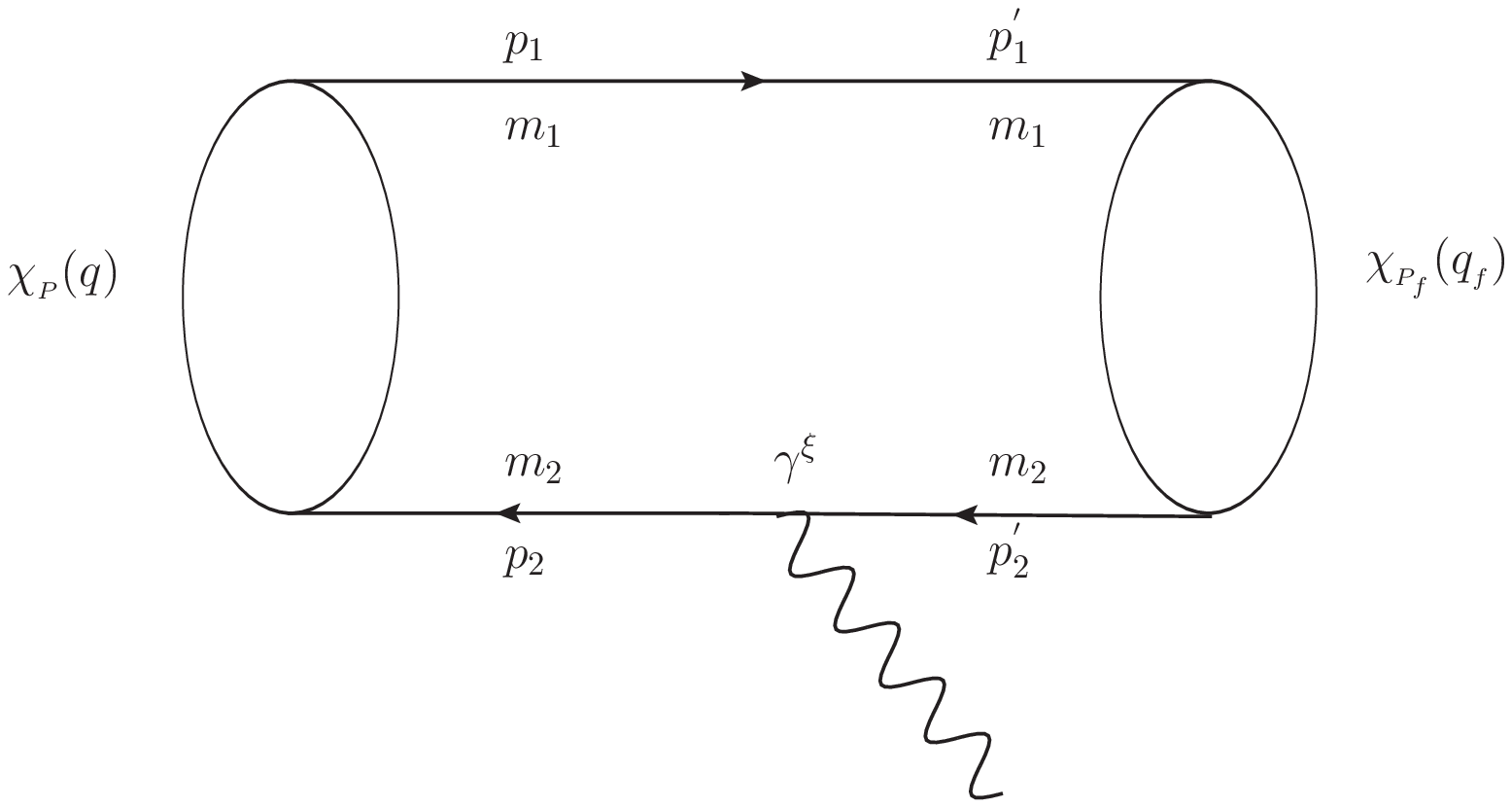}
\caption{{Feynman diagrams for the transition $X(3823)\rightarrow\chi_{_{cJ}}\gamma$. The two diagrams show that photons come from the quark and the anti-quark, respectively}.}
\end{center}
\end{figure}

Invariant amplitude ${\cal M}^{\xi}$ consists of two parts, corresponding to the two subgraphs in Figure 1, where photons are emitted from quark and anti-quark, respectively. The amplitude can be written as
\begin{eqnarray}
&&{\cal M}^{\xi}=\int\frac{d^{4}q}{(2\pi)^4}\frac{d^{4}q_{_f}}{(2\pi)^4}Tr[\bar{\chi}_{_{P_f}}(q_{_f})Q_{_1}e\gamma^{\xi}\chi_{_P}(q)(2\pi)^{4}\delta^{4}(p_{_2}-p'_{_2})S^{-1}_{_2}(-p_{_2})
\nonumber\\&&\hspace{1.3cm}+\bar{\chi}_{_{P_f}}(q_{_f})(2\pi)^{4}\delta^{4}(p_{_1}-p'_{_1})S^{-1}_{_1}(p_{_1})\chi_{_P}(q)Q_{_2}e\gamma^{\xi}],
\end{eqnarray}
where $\chi_{_P}(q)$, ${\chi}_{_{P_f}}(q_{_f})$ are the relativistic BS wave functions for $X(3823)$ and $\chi_{_{cJ}}$, respectively. $q$ and $q_{_f}$ are the internal relative momenta of the initial and final mesons, respectively. $p_{_1}$, $p_{_2}$, $p'_{_1}$ and $p'_{_2}$ are the momenta of quark and anti-quark in the initial and final mesons, respectively. $Q_{_1}$ and $Q_{_2}$ are the electric charges (in unit of $e$) of quark and anti-quark, respectively. $S_{_1}$, $S_{_2}$ are the propagators for quark and anti-quark.

Since instead of BS equation, the Salpeter equation is solved, where we have used the instantaneous approximation, we need to make the same approximation to the invariant amplitude. Here we only show the amplitude formula we used, interested reader can find the details in Ref. \cite{C.H.C J.K.Chen G.L.Wang2006}. {The amplitude has the following form
\begin{eqnarray}\label{allamp}
&&{\cal M}^{\xi}=\int{\frac{d^3q_{_\perp}}{(2\pi)^{3}}}Tr\bigg\{Q_{_1}e\frac{\slashed{P}}{M}\bigg[\bar{\varphi}^{++}_{_{f}}(q_{_\perp}+\alpha_{_2}P_{_{f_\perp}})\gamma^{\xi}\varphi^{++}_{_i}(q_{_\perp})+\bar{\varphi}^{++}_{_{f}}(q_{_\perp}+\alpha_{_2}P_{_{f_\perp}})\gamma^{\xi}\psi^{-+}_{_{1i}}(q_{_\perp})
\nonumber\\&&\hspace{2.5cm}-\bar{\psi}^{-+}_{_{1f}}(q_{_\perp}+\alpha_{_2}P_{_{f_\perp}})\gamma^{\xi}\varphi^{--}_{_i}(q_{_\perp})-\bar{\psi}^{+-}_{_{1f}}(q_{_\perp}+\alpha_{_2}P_{_{f_\perp}})\gamma^{\xi}\varphi^{++}_{_i}(q_{_\perp})
\nonumber\\&&\hspace{2.5cm}+\bar{\varphi}^{--}_{_{f}}(q_{_\perp}+\alpha_{_2}P_{_{f_\perp}})\gamma^{\xi}\psi^{+-}_{_{1i}}(q_{_\perp})+\bar{\varphi}^{--}_{_{f}}(q_{_\perp}+\alpha_{_2}P_{_{f_\perp}})\gamma^{\xi}\varphi^{--}_{_i}(q_{_\perp})\bigg]
\nonumber\\&&\hspace{2.5cm}+Q_{_2}e\bigg[\bar{\varphi}^{++}_{_f}(q_{_\perp}-\alpha_{_1}P_{_{f_\perp}})\frac{\slashed{P}}{M}\varphi^{++}_{_i}(q_{_\perp})+\bar{\varphi}^{++}_{_f}(q_{_\perp}-\alpha_{_1}P_{_{f_\perp}})\frac{\slashed{P}}{M}\psi^{+-}_{_{2i}}(q_{_\perp})
\nonumber\\&&\hspace{2.5cm}-\bar{\psi}^{+-}_{_{2f}}(q_{_\perp}-\alpha_{_1}P_{_{f_\perp}})\frac{\slashed{P}}{M}\varphi^{--}_{_i}(q_{_\perp})-\bar{\psi}^{-+}_{_{2f}}(q_{_\perp}-\alpha_{_1}P_{_{f_\perp}})\frac{\slashed{P}}{M}\varphi^{++}_{_i}(q_{_\perp})
\nonumber\\&&\hspace{2.5cm}+\bar{\varphi}^{--}_{_f}(q_{_\perp}-\alpha_{_1}P_{_{f_\perp}})\frac{\slashed{P}}{M}\psi^{-+}_{_{2i}}(q_{_\perp})-\bar{\varphi}^{--}_{_f}(q_{_\perp}-\alpha_{_1}P_{_{f_\perp}})\frac{\slashed{P}}{M}\varphi^{--}_{_i}(q_{_\perp})\bigg]\gamma^{\xi}\bigg\}.
\end{eqnarray}}
Where $M$ is the mass of $X(3823)$, $\alpha_{_1}=\frac{m_{_1}}{m_{_1}+m_{_2}}$ and $\alpha_{_2}=\frac{m_{_2}}{m_{_1}+m_{_2}}$ with the quark mass $m_{_1}=m_{_c}$ and anti-quark mass $m_{_2}=m_{_c}$. $\varphi^{++}_{_{i,f}}$ is the positive energy wave function, {$\varphi^{--}_{_{i,f}}$ is the negative energy wave function,} $i,f$ stand for initial and final states, respectively. $P_{_{f_\perp}}$ and $\bar{\varphi}^{++}$ are defined as $P^{\mu}_{_{f_\perp}}=P^{\mu}_{_f}-(P\cdot{P_{_f}/M^{2})P^{\mu}}$ and $\bar{\varphi}^{++}=\gamma_{_0}(\varphi^{++})^{\dagger}\gamma_{_0}$, respectively. {In order to compare these wave functions, we give their definitions in the initial state \cite{C.H.C J.K.Chen G.L.Wang2006}}
{\begin{eqnarray}\label{+-}
&&\varphi^{++}_{_i}\equiv\frac{\Lambda^{+}_{_1}(q_{_\perp})\eta_{_P}(q_{_\perp})\Lambda^{+}_{_2}(q_{_\perp})}{M-\omega_{_1}-\omega_{_{2}}},
~~~~~~~~~~~~\varphi^{--}_{_i}\equiv-\frac{\Lambda^{-}_{_1}(q_{_\perp})\eta_{_P}(q_{_\perp})\Lambda^{-}_{_2}(q_{_\perp})}{M+\omega_{_1}+\omega_{_{2}}},
\nonumber\\&&\psi^{-+}_{_{1i}}\equiv\frac{\Lambda^{-}_{_1}(q_{_\perp})\eta_{_P}(q_{_\perp})\Lambda^{+}_{_2}(q_{_\perp})}{M+\omega_{_1}+\omega_{_{1f}}-E_{_f}},
~~~~~~~~~~~~\psi^{+-}_{_{i}}\equiv\frac{\Lambda^{+}_{_1}(q_{_\perp})\eta_{_P}(q_{_\perp})\Lambda^{-}_{_2}(q_{_\perp})}{M-\omega_{_1}-\omega_{_{1f}}-E_{_f}},
\nonumber\\&&\psi^{-+}_{_{2i}}\equiv\frac{\Lambda^{-}_{_1}(q_{_\perp})\eta_{_P}(q_{_\perp})\Lambda^{+}_{_2}(q_{_\perp})}{M-\omega_{_2}-\omega_{_{2f}}-E_{_f}},
~~~~~~~~~~~~\psi^{+-}_{_{2i}}\equiv\frac{\Lambda^{+}_{_1}(q_{_\perp})\eta_{_P}(q_{_\perp})\Lambda^{-}_{_2}(q_{_\perp})}{M+\omega_{_2}+\omega_{_{2f}}-E_{_f}},
\end{eqnarray}}
{with}
{$
\omega_{_{i}}=\sqrt{m^2_{_i}-q^2_{_{i\perp}}}$, $\omega_{_{if}}=\sqrt{m^2_{_{if}}-q^2_{_{if_{\perp}}}}$ and $\Lambda^{\pm}_{_{i}}(q_{_\perp})=\frac{1}{2\omega_{_{i}}}[\frac{\slashed{P}}{M}\omega_{_{i}}\pm J(m_{_i}+\slashed{q}_{_\perp})]$,
where $i=1, 2$, $J=1$ for the quark ($i=1$) and $J=-1$ for the anti-quark ($i=2$). $E_{_f}$ is energy of the final meson, $\eta_{_{P}}(q_{_\perp})=\int\frac{dk_{_{\perp}}}{(2\pi)^3}V(q_{_\perp},k_{_\perp})\varphi_{_i}(k_{_\perp})$, where the Cornell potential $V(q_{_\perp},k_{_\perp})$ is chosen \cite{K.C.S AND G.L.Wang2004,liwei}.}

{As can be seen from the definition of Eq.(\ref{+-}), the numerators of these wave functions have the similar structure and the numerator values are comparable. But the denominator of $\varphi^{++}$, $M-\omega_{_1}-\omega_{_{2}}\sim 0$, is much smaller than others, for example the denominator of $\varphi^{--}$, $M+\omega_{_1}+\omega_{_{2}}\sim 2M$. So the contribution of $\varphi^{++}$ is much larger than others. Therefore, to simplify the calculation, the decay amplitude in Eq.(\ref{allamp}) can be written as
\begin{eqnarray}\label{amp}
&&{\cal M}^{\xi}=\int{\frac{d^3q_{_\perp}}{(2\pi)^{3}}}Tr[Q_{_1}e\frac{\slashed{P}}{M}\bar{\varphi}^{++}_{_f}(q_{_\perp}+\alpha_{_2}P_{_{f_\perp}})\gamma^{\xi}\varphi^{++}_i(q_{_\perp})
\nonumber\\&&\hspace{1.3cm}+Q_{_2}e~\bar{\varphi}^{++}_{_f}(q_{_\perp}-\alpha_{_1}P_{_{f_\perp}})\frac{\slashed{P}}{M}\varphi^{++}_{_i}(q_{_\perp})\gamma^{\xi}].
\end{eqnarray}
We will compare the decay widths given by Eq.(\ref{allamp}) and Eq.(\ref{amp}) in Sec. III to prove that the decay width formula retaining only the positive wave function $\varphi^{++}$ is simple and effective.}

\subsection{The Relativistic Wave Functions}
Though the BS equation is the relativistic dynamic equation, it can not provide us the form of a relativistic wave function for a bound state. In previous studies, the relativistic formula of the wave function for a meson with definite $J^{PC}$ numbers is constructed requiring each term in the function having the same $J^{PC}$ as the meson. With this wave function formula as input, the corresponding Salpeter equation is solved for different $J^{PC}$ state, for example see Ref. \cite{C-H.Chang.G.L.Wang2010}.

Here we do not show the detail how to solve the corresponding Salpeter equation, but only show the relativistic wave function of $X(3823)$ as a $2^{--}$ state \cite{T.H.Wang2016}
\begin{eqnarray}\label{2--}
&&\varphi_{_{2^{--}}}(q_{_\perp})=i\epsilon_{_{\mu\nu\alpha\beta}}
\frac{P^{\nu}}{M}q^{\alpha}_{_\perp}\epsilon^{\beta\delta}q_{_{\perp\delta}}\gamma^{\mu}
\left(f_{_1}+\frac{\slashed{P}}{M}f_{_2}
+\frac{\slashed{P}\slashed{q}_{_\perp}}{Mm_c}f_{_2}\right),
\end{eqnarray}
where $\epsilon^{\beta\delta}$ is the polarization tensor of $X(3823)$ and $\epsilon_{_{\mu\nu\alpha\beta}}$ is the Levi-Civita simbol. $f_{_1}$ and $f_{_2}$ are independent radial wave functions and they are function of $-q^2_{_\perp}$.

The positive energy wave function for a $2^{--}$ state is
\begin{eqnarray}\label{2--2}
&&\varphi^{++}_{_{2^{--}}}(q_{_\perp})=i\epsilon_{_{\mu\nu\alpha\beta}}\frac{P^{\nu}}{M}q^{\alpha}_{_\perp}q_{_{\perp\delta}}\epsilon^{\beta\delta}\gamma^{\mu}[F_{_1}+\frac{\slashed{P}}{M}F_{_2}
+\frac{\slashed{P}\slashed{q}_{_\perp}}{M^{2}}F_{_3}],
\end{eqnarray}
where
$$F_{_1}=\frac{1}{2}[f_{_1}-\frac{\omega_{_c}}{m_{_c}}f_{_2}],
~~F_{_2}=-\frac{1}{2}[\frac{m_{_c}}{\omega_{_c}}f_{_1}-f_{_2}],
~~F_{_3}=-\frac{M}{\omega_{_c}}F_{_1},$$
{where $\omega_{_{c}}=\sqrt{m^2_{_c}-q^2_{_\perp}}$ is the energy of charm quark.}
According to the method in Ref.\cite{wang2022}, we know that $F_1$ and $F_2$ terms are dominant $D$ partial waves which will survive in the non-relativistic limit, while the relativistic term including $F_3$ is $F$ partial wave.

The positive energy wave function for the $\eta_{_{c}}$ ($0^{-+}$) is written as \cite{K.C.S AND G.L.Wang2004}
\begin{eqnarray}\label{0-+}
&&\varphi^{++}_{_{0^{-+}}}(q_{_{f_{\perp}}})=[A_{_{f_1}}+\frac{\slashed{P}_{_{f_\perp}}}{M_{_f}}A_{_{f_2}}+\frac{\slashed{P}_{_f}\slashed{q}_{_{f_{\perp}}}}{M^{2}_{_f}}A_{_{f_3}}]\gamma^{5},
\end{eqnarray}
where $A_{_{f_1}}$ and $A_{_{f_2}}$ terms are dominant $S$ waves, relativistic $A_{_{f_3}}$ term is $P$ wave, with
$$
A_{_{f_1}}=\frac{M_{_f}}{2}[\frac{\omega_{_f}}{m_{_f}}a_{_1}+a_{_2}],
~~A_{_{f_2}}=\frac{M_{_f}}{2}[a_{_1}+\frac{m_{_f}}{\omega_{_f}}a_{_2}],
~~A_{_{f_3}}=-\frac{M_{_f}}{\omega_{_f}}A_{_{f_1}},
$$
{
$\omega_{_{f}}=\sqrt{m^2_{_{f}}-q^2_{_{f\perp}}},~m_{_f}=m_{_c}
$}, $a_{_1}$ and $a_{_2}$ are independent radial wave functions, and they are function of $-q^2_{_{f_{\perp}}}$.

The positive energy wave function for the $\chi_{_{c0}}$ ($0^{++}$) is written as \cite{G.L.Wang2007}
\begin{eqnarray}\label{0++}
&&\varphi^{++}_{_{0^{++}}}(q_{_{f_{\perp}}})=B_{_{f_1}}+\frac{\slashed{q}_{_{f_\perp}}}{M_{_f}}B_{_{f_2}}+\frac{\slashed{P}_{_f}\slashed{q}_{_{f_{\perp}}}}{M^{2}_{_f}}B_{_{f_3}},
\end{eqnarray}
where  $B_{_{f_2}}$ and $B_{_{f_3}}$ terms are dominant $P$ waves, relativistic $B_{_{f_1}}$ term is $S$ wave, with
$$B_{_{f_1}}=\frac{q^{2}_{_{f_\perp}}}{2m_{_f}}[b_{_1}+\frac{m_{_f}}{\omega_{_f}}b_{_2}],
~~B_{_{f_2}}=\frac{M_{_f}}{2}[b_{_1}+\frac{m_{_f}}{\omega_{_f}}b_{_2}],~~B_{_{f_3}}=\frac{M_{_f}}{2}[\frac{\omega_{_f}}{m_{_f}}b_{_1}+b_{_2}],
$$ $b_{_1}$ and $b_{_2}$ are independent radial wave functions.

The positive energy wave function for the $1^{++}$ state $\chi_{_{c1}}$ can be written as \cite{G.L.Wang2007}
\begin{eqnarray}\label{1++}
&&\varphi^{++}_{1^{++}}(q_{_{f_{\perp}}})=i\epsilon_{_{\mu\nu\alpha\beta}}\frac{P^{\nu}_{_f}}{M_{_f}}q^{\alpha}_{_{f_{\perp}}}\epsilon^{\beta}_{_f}\gamma^{\mu}[C_{_{f_1}}+\frac{\slashed{P}_{_f}}{M_{_f}}C_{_{f_2}}
+\frac{\slashed{P}_{_f}\slashed{q}_{_{f_\perp}}}{M^{2}_{_f}}C_{_{f_3}}],
\end{eqnarray}
where $C_{_{f_1}}$ and $C_{_{f_2}}$ terms are dominant $P$ waves, relativistic $C_{_{f_3}}$ term is $D$ wave, with
$$C_{_{f_1}}=\frac{1}{2}[c_{_1}+\frac{\omega_{_f}}{m_{_f}}c_{_2}],
~~C_{_{f_2}}=-\frac{1}{2}[\frac{m_{_f}}{\omega_{_f}}c_{_1}+c_{_2}],
~~C_{_{f_3}}=-\frac{M_{_f}}{\omega_{_f}}C_{_{f_1}},
$$ $c_{_1}$ and $c_{_2}$ are independent radial wave functions.

The positive energy part of wave function for $2^{++}$ state $\chi_{_{c2}}$ can be written as \cite{G-L Wang2009}
\begin{eqnarray}\label{2++}
&&\varphi^{++}_{_{2^{++}}}(q_{_{f_{\perp}}})=\epsilon_{_{f,\mu\nu}}q^{\mu}_{_{f_{\perp}}}q^{\nu}_{_{f_{\perp}}}[D_{_{f_1}}+\frac{\slashed{P}_{_f}}{M_{_f}}D_{_{f_2}}+\frac{\slashed{q}_{_{f_{\perp}}}}{M_{_f}}D_{_{f_3}}+\frac{\slashed{P}_{_f}\slashed{q}_{_{f_{\perp}}}}{M^{2}_{_f}}D_{_{f_4}}]
\nonumber\\&&\hspace{2.0cm}+M_{_f}\epsilon_{_{f,\mu\nu}}\gamma^{\mu}q^{\nu}_{_{f_{\perp}}}[D_{_{f_5}}+\frac{\slashed{P}_{_f}}{M_{_f}}D_{_{f_6}}
+\frac{\slashed{P}_{_f}\slashed{q}_{_{f_{\perp}}}}{M^{2}_{_f}}D_{_{f_7}}],
\end{eqnarray}
where $D_{_{f_5}}$ and $D_{_{f_6}}$ terms are $P$ partial waves, $D_{_{f_1}}$, $D_{_{f_2}}$ and $D_{_{f_7}}$ terms are $D$ partial waves, while $D_{_{f_3}}$ and $D_{_{f_4}}$ terms are $F$ partial waves, with
$$D_{_{f_1}}=\frac{1}{2M_{_f}m_{_f}\omega_{_f}}[\omega_{_f}q^{2}_{_{f_{\perp}}}d_{_3}+m_{_f}q^{2}_{_{f_{\perp}}}d_{_4}+M^{2}_{_f}\omega_{_f}d_{_5}-M^{2}_{_f}m_{_f}d_{_6}],$$
$$~~~~D_{_{f_2}}=\frac{M_{_f}}{2m_{_f}\omega_{_f}}[m_{_f}d_{_5}-\omega_{_f}d_{_6}],
~~~~D_{_{f_3}}=\frac{1}{2}[d_{_3}+\frac{m_{_f}}{\omega_{_f}}d_{_4}-\frac{M^{2}_{_f}}{m_{_f}\omega_{_f}}d_{_6}],$$
$$~~~~D_{_{f_4}}=\frac{1}{2}[\frac{\omega_{_f}}{m_{_f}}d_{_3}+d_{_4}-\frac{M^{2}_{_f}}{m_{_f}\omega_{_f}}d_{_5}],
~~~~D_{_{f_5}}=\frac{1}{2}[d_{_5}-\frac{\omega_{_f}}{m_{_f}}d_{_6}],$$
$$~~~~D_{_{f_6}}=\frac{1}{2}[-\frac{m_{_f}}{\omega_{_f}}d_{_5}+d_{_6}],
~~~~D_{_{f_7}}=\frac{M_{_f}}{2\omega_{_f}}[-d_{_5}+\frac{\omega_{_f}}{m_{_f}}d_{_6}],
$$ $d_i$ are independent radial wave functions. $2^{++}$ states are very complicated, there are two typical kinds of states, one is $P$ wave dominant state with small amount of $D$ and $F$ waves, the other is $F$ wave dominant state but with sizable components of $P$ and $D$ waves \cite{wang2022}.

For latter use, we show the non-relativistic forms of the wave functions. We know that in the non-relativistic limit, only the lowest order $q_{_{\perp}}$ (or  $q_{_{f_{\perp}}}$) term in wave function has contribution, and the wave function of each state contains only one independent radial wave function.
Considering the whole wave functions, the non-relativistic ones for $2^{--}$, $0^{-+}$, $0^{++}$, $1^{++}$ and $2^{++}$ states can be written as
\begin{eqnarray}\label{2--2}
&&\varphi^{++}_{_{2^{--}}}(q_{_\perp})=i\epsilon_{_{\mu\nu\alpha\beta}}\frac{P^{\nu}}{M}q^{\alpha}_{_\perp}q_{_{\perp\delta}}\epsilon^{\beta\delta}\gamma^{\mu}\left(1-\frac{\slashed{P}}{M}\right)F_{_1},
\end{eqnarray}
\begin{eqnarray}\label{0-+}
&&\varphi^{++}_{_{0^{-+}}}(q_{_{f_{\perp}}})=\left(1+\frac{\slashed{P}_{_{f_\perp}}}{M_{_f}}\right)\gamma^{5}A_{_{f_1}},
\end{eqnarray}
\begin{eqnarray}\label{0++}
&&\varphi^{++}_{_{0^{++}}}(q_{_{f_{\perp}}})=\left(\frac{\slashed{q}_{_{f_\perp}}}{M_{_f}}+\frac{\slashed{P}_{_f}\slashed{q}_{_{f_{\perp}}}}{M^{2}_{_f}}\right)B_{_{f_2}},
\end{eqnarray}
\begin{eqnarray}\label{1++}
&&\varphi^{++}_{1^{++}}(q_{_{f_{\perp}}})=i\epsilon_{_{\mu\nu\alpha\beta}}\frac{P^{\nu}_{_f}}{M_{_f}}q^{\alpha}_{_{f_{\perp}}}\epsilon^{\beta}_{_f}\gamma^{\mu}\left(1-\frac{\slashed{P}_{_f}}{M_{_f}}\right)C_{_{f_1}},
\end{eqnarray}
\begin{eqnarray}\label{2++}
&&\varphi^{++}_{_{2^{++}}}(q_{_{f_{\perp}}})=M_{_f}\epsilon_{_{f,\mu\nu}}\gamma^{\mu}q^{\nu}_{_{f_{\perp}}}\left(1-\frac{\slashed{P}_{_f}}{M_{_f}}\right)D_{_{f_5}}.
\end{eqnarray}

\subsection{The Form Factors}

Using Eq.(\ref{amp}), where we integrate internal ${q}_{_\perp}$ over the initial and final state wave functions, then obtain the amplitude described using form factors.\\
(1) For the channel $X(3823)$ $\rightarrow$ $\eta_{_c}(^1S_{_0})\gamma$, there are two form factors $h_{_1}$ and  $h_{_2}$,
\begin{eqnarray}
&&{\cal M}^{\xi}_{_1}=P^{\xi}\epsilon_{_{\mu\nu}}P^{\mu}_{_f}P^{\nu}_{_f}h_{_1}+\epsilon^{\xi}_{_\mu}P^{\mu}_{_f}h_{_2}.
\end{eqnarray}
(2) For $X(3823)$ $\rightarrow$ $\chi_{_{c0}}(^3P_{_0})\gamma$, there is only one form factor $t_{_1}$,
\begin{eqnarray}
&&{\cal M}^{\xi}_{_2}=i\epsilon^{\beta\xi\mu\nu}\epsilon_{_{\beta\alpha}}P_{_\mu}P_{_{f,\nu}}P^{\alpha}_{_{f}}t_{_1}.
\end{eqnarray}
(3) For $X(3823)$ $\rightarrow$ $\chi_{_{c1}}(^3P_{_1})\gamma$, there are five form factors $s_{_i}$,
\begin{eqnarray}
&&{\cal M}^{\xi}_{_3}=\epsilon_{_{\mu\nu}}P^{\xi}P^{\mu}_{_f}P^{\nu}_{_f}{P}\cdot\epsilon_{_f}s_{_1}
+\epsilon^{\xi}_{_\nu}P^{\nu}_{_f}{P}\cdot\epsilon_{_f}s_{_2}+\epsilon_{_{\mu\nu}}P^\xi\epsilon^{\mu}_{_f}P^{\nu}_{_f}s_{_3}
\nonumber\\&&\hspace{1.5cm}+\epsilon_{_{\mu\nu}}\epsilon^{\xi}_{_f}P^{\mu}_{_f}P^{\nu}_{_f}s_{_4}+\epsilon^{\xi}_{_\mu}\epsilon^{\mu}_{_f}s_{_5},
\end{eqnarray}
where $\epsilon_{_f}^{_\mu}$ is the polarization vector of $\chi_{_{c1}}$.\\
(4) For $X(3823)$ $\rightarrow$ $\chi_{_{c2}}(^3P_{_2})\gamma$ or $X(3823)$ $\rightarrow$ $\chi_{_{c2}}(^3F_{_2})\gamma$, the amplitude is more complicated, which can be represented by eight form factors $g_{_i}$,
\begin{eqnarray}
&&{\cal M}^{\xi}_{_4}=i\epsilon^{\beta\lambda{P_{_f}P}}\left(\epsilon_{_{\beta{P_{_f}}}}\epsilon_{_{f,\lambda{P}}}P^{\xi}g_{_1}
+\epsilon^{\phi}_{_{\beta}}\epsilon_{_{f,\lambda\phi}}P^{\xi}g_{_2}
+\epsilon^{\xi}_{_\beta}\epsilon_{_{f,\lambda{P}}}g_{_3}
+\epsilon_{_{\beta{P_{_f}}}}\epsilon^{\xi}_{_{f,\lambda}}g_{_4}\right)
\nonumber\\&&\hspace{1cm}+i\epsilon^{\beta\xi{P_{_f}P}}\left(\epsilon_{_{\beta{P_{_f}}}}\epsilon_{_{f,PP}}g_{_5}
+\epsilon^{\phi}_{_{\beta}}\epsilon_{_{f,\phi{P}}}g_{_6}\right)
+i\epsilon^{\beta\xi\lambda{P}}\left(\epsilon_{_{\beta{P_{_f}}}}\epsilon_{_{f,\lambda{P}}}g_{_7}
+\epsilon^{\phi}_{_{\beta}}\epsilon_{_{f,\lambda\phi}}g_{_8}\right),
\end{eqnarray}
where $\epsilon_{_{f,\mu\nu}}$ is the polarization tensor of $\chi_{_{c2}}(^3P_{_2})$, and we have used some abbreviations, for example, $\epsilon^{\beta\lambda{P_{_f}P}}\epsilon_{_{\beta{P_{_f}}}}\epsilon_{_{f,\lambda{P}}}
=\epsilon^{\beta\lambda{\mu\nu}}P_{_{f,\mu}}P_{_{\nu}}\epsilon_{_{\beta\alpha}}{P^{\alpha}_{_f}}\epsilon_{_{f,\lambda\rho}}{P}^{\rho}$.
If the final state is $^3F_{_2}$ state, the definitions of the form factors are same as those for $^3P_{_2}$ state. {Since the expressions of $h_{_i}$, $s_{_i}$, $t_{_i}$ and $g_{_i}$ are complex and long, their specific expressions are not given here, we put their detailed description in Appendix B.}

The thing to note here is that most of these form factors are not independent. Due to the Ward identity $(P_{\xi}-P_{f,\xi}){\cal M}^{\xi}_i=0$ ($i=1,2,3,4$), they are linked by the following constrain conditions:
\begin{eqnarray}
 h_{_2}=(M^2-ME_{_f})h_{_1},
\end{eqnarray}
\begin{eqnarray}
 s_{_2}=(M^2-ME_{_f})s_{_1}+s_{_4},~~ s_{_5}=(M^2-ME_{_f})s_{_3},
\end{eqnarray}
\begin{eqnarray}
 g_{_3}=(M^2-ME_{_f})g_{_1}+g_{_4}+g_{_7},~~ g_{_8}=-(M^2-ME_{_f})g_{_2}.
\end{eqnarray}
Other form factors such as $t_{_1}$, $g_{_5}$ and $g_{_6}$ are independent and have no such constraints.

{Then, the amplitude square for the EM decay of $X(3823)$ is}
{\begin{eqnarray}
\overline{{|\cal{M}|}^2}=\frac{1}{2J+1}\sum_{\gamma}{\varepsilon}^{(\gamma)}_{_\xi}{\varepsilon}^{(\gamma)}_{_{\xi^{'}}}{\cal M}^{\xi}{\cal M}^{\xi^{'}},
\end{eqnarray}}
where, $\varepsilon^{(\gamma)}_{_\xi}$ is the polarization vector of the final state photon $\gamma$, $J$ is the total angular momentum of the initial state. For the $X(3823)$ $\rightarrow$ $\eta_{_c}(^1S_{_0})\gamma$ decay channel, we have
\begin{eqnarray}
&&\overline{{|\cal{M}|}^2_1}=\frac{1}{5}\frac{4e^2}{9}h^2_{_1}M^2|\vec{P_{_f}}|^{4}.
\end{eqnarray}
For $X(3823)$ $\rightarrow$ $\chi_{_{c0}}(^3P_{_0})\gamma$
\begin{eqnarray}
&&\overline{{|\cal{M}|}^2_2}=\frac{1}{5}\frac{4e^2}{9}t^2_{_1}M^2|\vec{P_{_f}}|^{4}.
\end{eqnarray}
For $X(3823)$ $\rightarrow$ $\chi_{_{c1}}(^3P_{_1})\gamma$ and $X(3823)$ $\rightarrow$ $\chi_{_{c2}}(^3P_{_2})\gamma$, the modulus square of amplitudes is more complex, and for brevity they are placed in Appendix C for the reader's reference.

{Finally, the two-body decay width formulation can be written as
\begin{eqnarray}
\Gamma=\frac{|\vec{P_{_f}}|}{8{\pi}M^2}\overline{{|\cal{M}|}^2},
\end{eqnarray}
where, $|\vec{P_{_f}}|=\left(M^2-M_{_{f}}^2\right)/2M$.}

\subsection{Decay Widths in Non-relativistic Approximation}

Although this article presents a relativistic calculation, we like to give the decay width in the non-relativistic approximation, since the later has simplified formula and may help to see the problem clearly. Using the non-relativistic wave functions in Eqs.(12-16), we obtain the radiative decay widths of $X(3823)$.

For the $X(3823)$ $\rightarrow$ $\eta_{_c}(^1S_{_0})\gamma$ decay channel, we have
\begin{eqnarray}\label{eq28}
&&\Gamma_{1}=\frac{2\alpha E_{\gamma}^3}{9MM_{_{f}}}\bigg[\int\frac{q^2dqd cos \theta}{(2\pi)^2}\bigg(\frac{2q^2F_{_1}}{\sqrt{5M}}\bigg)\cdot1\cdot\bigg(\frac{(A_{_{f_1}}+A^{'}_{_{f_1}})}{\sqrt{M_{_f}}}\bigg)(3\cos^2\theta-1)\bigg]^2_{M_1},
\end{eqnarray}
where $E_{\gamma}$ is the energy of emitted photon, $q\equiv{\mid \vec q\mid}$, $\theta$ is the angle between $\vec{q}$ and $\vec{P_{_f}}$. In non-relativistic limit, since $\omega_c=m_c$, wave functions $F_{_1}$ and $A_{_{f_1}}$ are related to the original radial wave functions directly, $F_{_1}=F_{_1}(q)=f_{_1}\simeq-f_{_2}$, $A_{_{f_1}}=M_{_f}a_{_1}\simeq M_{_f}a_{_2}$. $A_{_{f_1}}(q_{_{f_{\perp}}})$ and $A^{'}_{_{f_1}}(q_{_{f_{\perp}}})$ correspond to the two diagrams in Fig.1, where the photons emitted by quark and anti-quark, respectively.
$A_{_{f_1}}(q_{_{f_{\perp}}})=A_{_{f_1}}(\sqrt{{\vec q}^2+2\alpha_{_{2}}{\vec q}\cdot{\vec P_{_{f}} }+\alpha^2_{_{2}}{\vec P_{_{f}}}^2})$ (where $q_{_{f_{\perp}}}=q_{_\perp}+\alpha_{_2}P_{_{f_\perp}}$) and $A^{'}_{_{f_1}}(q_{_{f_{\perp}}})=A^{'}_{_{f_1}}(\sqrt{{\vec q}^2-2\alpha_{_{1}}{\vec q}\cdot{\vec P_{_{f}} }+\alpha^2_{_{1}}{\vec P_{_{f}}}^2})$ (where $q_{_{f_{\perp}}}=q_{_\perp}-\alpha_{_1}P_{_{f_\perp}}$). Then it can be seen that we have already consider the recoil effect in the transition.

In the above equation of the decay width, the representations of the radial wave functions $\bigg(\frac{2q^2F_{_1}}{\sqrt{5M}}\bigg)$ and $\bigg(\frac{(A_{_{f_1}}+A^{'}_{_{f_1}})}{\sqrt{M_{_f}}}\bigg)$ are based on their normalization conditions,  $\int\frac{d^3q}{(2\pi)^3}\frac{4q^4F^2_{_1}(q)}{{5M}}=1$ for $2^{--}$ state and $\int\frac{d^3q_{_f}}{(2\pi)^3}\frac{4A^2_{_{f_1}}(q_{_f})}{M_{_f}}=1$ for   $0^{-+}$ state. Therefore, it can be seen from the formula of decay width that this is a $M_1$ magnetic radiative transition, and a subscript $M_1$ is marked.

For $X(3823)$ $\rightarrow$ $\chi_{_{c1}}(^3P_{_0})\gamma$, we have
\begin{eqnarray}\label{eq29}
&&\Gamma_{2}=\frac{2\alpha E_{\gamma}^3}{9MM_{_{f}}}\bigg[-\int\frac{q^2dqd cos \theta}{(2\pi)^2}\bigg(\frac{2q^2F_{_1}}{\sqrt{5M}}\bigg)\cdot1\cdot\bigg(\frac{q(B_{_{f_2}}-B^{'}_{_{f_2}})}{\sqrt{M^3_{_f}}}\bigg)
(\cos^3\theta-\cos\theta)\bigg]^2_{M_2},
\end{eqnarray}
where, subscript $M_2$ denote the $M_{2}$ magnetic radiative transitions. Normalization condition  $\int\frac{d^3q_{_f}}{(2\pi)^3}\frac{4q_{_f}^2B^2_{_{f_2}}(q_{_f})}{{M^3_{_f}}}=1$ for $0^{++}$ state has been considered.

For $X(3823)$ $\rightarrow$ $\chi_{_{c1}}(^3P_{_1})\gamma$, we have
\begin{eqnarray}\label{eq33}
&&\Gamma_{3}=\frac{7\alpha E_{\gamma}^3E^2_{_f}(M+M_{_f})^2}{36MM^3_{_{f}}}\bigg[\langle1\rangle^2_{_{E_1}}+\frac{16M(E_{_f}-E_{\gamma})}{7M_{_{f}}E_{_f}(M+M_{_f})}
\langle1\rangle_{_{E_1}}\langle2\rangle_{_{M_2}}
\nonumber\\&&\hspace{1cm}+\frac{4M_{_f}}{E_{_{f}}(M+M_{_f})}\cdot\langle1\rangle_{_{E_1}}\langle3\rangle_{_{M_2}}\bigg],
\end{eqnarray}
where
\begin{eqnarray}
&&\langle1\rangle_{_{E_1}}=\int\frac{d^3q}{(2\pi)^3}\bigg(\frac{2q^2F_{_1}}{\sqrt{5M}}\bigg)\cdot\frac{1}{q}\cdot\bigg(\frac{\sqrt{2}q(C_{_{f_1}}+C^{'}_{_{f_1}})}{\sqrt{3M_{_f}}}\bigg)(3\cos^2\theta-1),
\end{eqnarray}
\begin{eqnarray}
&&\langle2\rangle_{_{M_2}}=\int\frac{d^3q}{(2\pi)^3}\bigg(\frac{2q^2F_{_1}}{\sqrt{5M}}\bigg)\cdot1\cdot\bigg(\frac{\sqrt{2}q(C_{_{f_1}}-C^{'}_{_{f_1}})}{\sqrt{3M_{_f}}}\bigg)(3\cos^3\theta-\cos\theta),
\end{eqnarray}
\begin{eqnarray}
&&\langle3\rangle_{_{M_2}}=\int\frac{d^3q}{(2\pi)^3}\bigg(\frac{2q^2F_{_1}}{\sqrt{5M}}\bigg)\cdot1\cdot\bigg(\frac{\sqrt{2}q(C_{_{f_1}}-C^{'}_{_{f_1}})}{\sqrt{3M_{_f}}}\bigg)\bigg[(\frac{M+E_{_f}}{E_{\gamma}}+\frac{3M}{2M_{_f}}
\nonumber\\&&\hspace{1cm}-1)\cos^3\theta+(1-\frac{M}{M_{_f}})\cos\theta\bigg].
\end{eqnarray}
When giving the upper representation, the normalization condition for the $1^{++}$ state, $\int\frac{d^3q_{_f}}{(2\pi)^3}\frac{8q^2_{_f}C^2_{_{f_1}}(q_{_f})}{{3M_{_f}}}=1$ has been concerned.

For $X(3823)$ $\rightarrow$ $\chi_{_{c2}}(^3P_{_2})\gamma$, we have
\begin{eqnarray}\label{eq37}
&&\Gamma_{4}=\frac{7\alpha E_{\gamma}^3E^2_{_f}(M+M_{_f})^2}{36MM^3_{_{f}}}\bigg[\bigg(1+\frac{4E_{\gamma}^2}{7(M+M_{_f})^2}\bigg)\langle4\rangle^2_{_{E_1}}+
\frac{4}{E_{\gamma}}\bigg(1+\frac{4E_{\gamma}^2}{7(M+M_{_f})^2}\bigg)\langle4\rangle_{_{E_1}}\langle5\rangle_{_{M_2}}
\nonumber\\&&\hspace{1cm}+\frac{4}{7E_{_{f}}(M+M_{_f})^2}(-8E_{_f}E_{\gamma}+2M_{_f}E_{\gamma}-7M^2+3MM_{_f}+10M^2_{_f})\langle4\rangle_{_{E_1}}\langle6\rangle_{_{M_2}}\bigg],
\end{eqnarray}
where
\begin{eqnarray}
&&\langle4\rangle_{_{E_1}}=\int\frac{d^3q}{(2\pi)^3}\bigg(\frac{2q^2F_{_1}}{\sqrt{5M}}\bigg)\cdot\frac{1}{q}\cdot\bigg(\frac{\sqrt{M_{_f}}q(D_{_{f_5}}+D^{'}_{_{f_5}})}{\sqrt{3}}\bigg)(3\cos^2\theta-1),
\end{eqnarray}
\begin{eqnarray}
&&\langle5\rangle_{_{M_2}}=\int\frac{d^3q}{(2\pi)^3}\bigg(\frac{2q^2F_{_1}}{\sqrt{5M}}\bigg)\cdot1\cdot\bigg(\frac{\sqrt{M_{_f}}q(D_{_{f_5}}-D^{'}_{_{f_5}})}{\sqrt{3}}\bigg)(5\cos^3\theta-3\cos\theta),
\end{eqnarray}
\begin{eqnarray}
&&\langle6\rangle_{_{M_2}}=\int\frac{d^3q}{(2\pi)^3}\bigg(\frac{2q^2F_{_1}}{\sqrt{5M}}\bigg)\cdot1\cdot\bigg(\frac{\sqrt{M_{_f}}q(D_{_{f_5}}-D^{'}_{_{f_5}})}{\sqrt{3}}\bigg)(\cos^3\theta-\cos\theta).
\end{eqnarray}
Here, $\int\frac{d^3q_{_f}}{(2\pi)^3}\frac{4q^2_{_f}{M_{_f}}D^2_{_{f_5}}}{{3}}=1$ is the normalization condition of $2^{++}$ state.

The non-relativistic expression of decay widths in Eqs.(\ref{eq28},\ref{eq29},\ref{eq33},\ref{eq37}) can be further simplified. Since in radiative decay, compared with initial meson mass $M$, the recoil momentum $|\vec {P_f}|=E_{\gamma}\equiv r$ is usually a small quantity, for example, in the radiative decays of $X(3823)$ to $\eta_{_c}$, $\chi_{_{c0}}$, $\chi_{_{c1}}$ and $\chi_{_{c2}}$, the recoil momenta are $0.746$ MeV, $0.386$ MeV, $0.298$ MeV and $0.256$ MeV, respectively.
Then the wave functions, for example, $A_{_{f_1}}\left(\sqrt{{ q}^2+q r \cos\theta+0.25 r^2}\right)=A_{_{f_1}}(\frac{r}{M} M \cos\theta)$  and $A^{'}_{_{f_1}}(-\frac{r}{M} M \cos\theta)$, can be expanded in a dimensionless quantity $\frac{r}{M}\cos\theta$. If the first four terms of Taylor expansion are retained, then we have
\begin{eqnarray}
&&A_{_{f_1}}(\frac{r}{M}\cos\theta)=A_{_{f_1}}+\frac{\partial A_{_{f_1}}}{\partial(\frac{r}{M}\cos\theta)}(\frac{r}{M}\cos\theta)+\frac{1}{2!}\frac{\partial^2 A_{_{f_1}}}{\partial(\frac{r}{M}\cos\theta)^2}(\frac{r}{M}\cos\theta)^2+
\nonumber\\&&\hspace{3cm}\frac{1}{3!}\frac{\partial^3 A_{_{f_1}}}{\partial(\frac{r}{M}\cos\theta)^3}(\frac{r}{M}\cos\theta)^3,\nonumber
\end{eqnarray}
\begin{eqnarray}
&&A^{'}_{_{f_1}}(-\frac{r}{M}\cos\theta)=A_{_{f_1}}-\frac{\partial A_{_{f_1}}}{\partial(\frac{r}{M}\cos\theta)}(\frac{r}{M}\cos\theta)+\frac{1}{2!}\frac{\partial^2 A_{_{f_1}}}{\partial(\frac{r}{M}\cos\theta)^2}(\frac{r}{M}\cos\theta)^2-
\nonumber\\&&\hspace{3cm}\frac{1}{3!}\frac{\partial^3 A_{_{f_1}}}{\partial(\frac{r}{M}\cos\theta)^3}(\frac{r}{M}\cos\theta)^3.\nonumber
\end{eqnarray}
So only even power of $r$ exists in
$$A_{_{f_1}}+A^{'}_{_{f_1}}=2 A_{_{f_1}}+\frac{r^2}{M^2}\cos^2\theta~\frac{\partial^2 A_{_{f_1}}}{\partial(\frac{r}{M}\cos\theta)^2},$$
and odd power of $r$ exists in
$$A_{_{f_1}}-A^{'}_{_{f_1}}=2~\frac{r}{M}\cos\theta~\frac{\partial A_{_{f_1}}}{\partial(\frac{r}{M}\cos\theta)}+
\frac{1}{3}\frac{r^3}{M^3}\cos^3\theta~\frac{\partial^3 A_{_{f_1}}}{\partial(\frac{r}{M}\cos\theta)^3}.$$ Then after integrating the angle $\theta$, the lowest order contribution in decay width $\Gamma_1$ for $X(3823)$ $\rightarrow$ $\eta_{_c}(^1S_{_0})\gamma$ is
\begin{eqnarray}
&&\Gamma_{1}=\frac{32\alpha r^7}{10125\pi^4M^6M^2_{_{f}}}\bigg(\int dqq^4F_{_1}\frac{\partial^2A_{_{f_1}}}{\partial(\frac{r}{M}\cos\theta)^2}\bigg)^2_{M_1},
\end{eqnarray}
where we can see that the leading order $2 A_{f_1}$ does not contribute, which is consistent with the non-relativistic results in Refs.\cite{T.B2005,rosner}.

For $X(3823)$ $\rightarrow$ $\chi_{_{c1}}(^3P_{_0})\gamma$, we find that the contribution of $E_1$ transition expanded to all orders is zero, which also confirms the results in Refs.\cite{T.B2005,rosner}. Further, only the $M_2$ transition has contribution, and the lowest order result is
\begin{eqnarray}
&&\Gamma_{2}=\frac{32\alpha r^5}{10125\pi^4M^4M^4_{_{f}}}\bigg(\int dq q^5 F_{_1}\frac{\partial B_{_{f_2}}}{\partial(\frac{r}{M}\cos\theta)}\bigg)^2_{M_2}.
\end{eqnarray}

The decay widths of $X(3823)$ $\rightarrow$ $\chi_{_{c1}}(^3P_{_1})\gamma$ and $X(3823)$ $\rightarrow$ $\chi_{_{c2}}(^3P_{_2})\gamma$ can be simplified as
\begin{eqnarray}
&&\Gamma_{3}=\frac{56\alpha r^5(M+M_{_f})^2}{10125\pi^4M^4M^2_{_{f}}}\bigg[\frac{r^2}{3M^2}
\bigg(\int dq q^4 F_{_1} \frac{\partial^2C_{_{f_1}}}{\partial(\frac{r}{M}\cos\theta)^2}\bigg)^2_{E_1}
\nonumber\\&&\hspace{1cm}+2\bigg(\int dq q^4 F_{_1}\frac{\partial^2C_{_{f_1}}}{\partial(\frac{r}{M}\cos\theta)^2}\bigg)_{E_1}
 \bigg(\int dq q^{5}F_{_1}\frac{\partial C_{_{f_1}}}{\partial(\frac{r}{M}\cos\theta)}\bigg)_{M_2}\bigg],
\end{eqnarray}
\begin{eqnarray}
&&\Gamma_{4}=\frac{28\alpha r^5(M+M_{_f})^2}{10125\pi^4M^4}\bigg[\frac{r^2}{3M^2}
\bigg(\int dq q^4 F_{_1} \frac{\partial^2D_{_{f_5}}}{\partial(\frac{r}{M}\cos\theta)^2}\bigg)^2_{E_1}
\nonumber\\&&\hspace{1cm}-\frac{2r}{7M}\bigg(\int dq q^4 F_{_1}\frac{\partial^2D_{_{f_5}}}{\partial(\frac{r}{M}\cos\theta)^2}\bigg)_{E_1}
 \bigg(\int dq q^{5}F_{_1}\frac{\partial D_{_{f_5}}}{\partial(\frac{r}{M}\cos\theta)}\bigg)_{M_2}\bigg],
\end{eqnarray}
where we retain the lowest order contribution of the $E_1$ transition and the lowest cross term between $E_1$ and $M_2$.

From the simplified expression of non-relativistic decay widths, it can be seen that, $X(3823)$ $\rightarrow$ $\eta_{_c}(^1S_{_0})\gamma$ decay is a $M_1$ transition. For $X(3823)$ $\rightarrow$ $\chi_{_{c1}}(^3P_{_0})\gamma$, the $E_1$ transition has zero contribution, then its contrition comes from the $M_2$ transition. While the main contributions of $X(3823)$ $\rightarrow$ $\chi_{_{c1}}(^3P_{_1})\gamma$ and $X(3823)$ $\rightarrow$ $\chi_{_{c2}}(^3P_{_2})\gamma$ come from the $E_1$ transition, so we conclude that the decay widths of $X(3823)$ $\rightarrow$ $\chi_{_{c1}}(^3P_{_1})\gamma$ and $X(3823)$ $\rightarrow$ $\chi_{_{c2}}(^3P_{_2})\gamma$ are much larger than those of $X(3823)$ $\rightarrow$ $\eta_{_c}(^1S_{_0})\gamma$ and $X(3823)$ $\rightarrow$ $\chi_{_{c1}}(^3P_{_0})\gamma$.

\section{RESULTS AND DISCUSSIONS}
\subsection{Masses}
In our calculation, some model-dependent parameters have been used, for example, the mass of the charm quark is fixed at $m_{_c}=1.62$ GeV \cite{G.L.Wang.T.F.Feng.X.G.Wu2020}.
Since $V_{_0}$ in the kernel originates from QCD non-perturbative effects, its value is to account the states with $J^{PC}$, so we fix it by fitting the masses of the ground states. Thus the parameter $V_{_0}$ vary with $J^{PC}$. And we vary the free parameter $V_{_0}$ \cite{C-H.Chang.G.L.Wang2010} to fit the mass of the ground state. For example, $M_{^{3}D_{_2}(1D)}=3.823$ GeV \cite{R. L. Workman2022} is actually not our prediction, but an input, while those of the first and second radial excited states are our predictions,
\begin{equation}
M_{^{3}D_{_2}(2D)}=4.154~ \rm{GeV},~~M_{^{3}D_{_2}(3D)}=4.408~ \rm{GeV}.
\end{equation}
{For other charmonia, we have calculated the mass spectrum in Ref. \cite{C-H.Chang.G.L.Wang2010}. For example, the masses of some highly excited states are predicted as,} $$M_{\eta_{_{c}}(3S)}=3.949~ \rm{GeV}, ~~~M_{\chi_{_{c2}}(1F)}=4.038~ \rm{GeV},~~~M_{\chi_{_{c2}}(2F)}=4.314 \rm{GeV},$$
$$M_{\chi_{_{c0}}(3P)}=4.140~ \rm{GeV},~~~M_{\chi_{_{c1}}(3P)}=4.229~ \rm{GeV}, ~~~M_{\chi_{_{c2}}(3P)}=4.271~ \rm{GeV}.$$

{It can be seen from Ref.\cite{C-H.Chang.G.L.Wang2010}, most of our predictions about the mass spectrum consist well with experimental data, especially the case of bottomonium. However, there are still some states whose theoretical masses are different from the experimental data. For example, our prediction of $M_{\chi_{_{c1}}(2P)}=3.929$ GeV \cite{C-H.Chang.G.L.Wang2010}, while the data is $M_{X(3872)}=3.872$ GeV, another is the mass of $\eta_c(2S)$, our prediction $3.576$ GeV is lower than data $3.636$ GeV. To see the difference in decays, for these two states, we use the theoretical mass as well as the experimental data to calculate the decay width, and give two groups of results.}

\subsection{Wave functions}
\begin{figure}
\centering
\includegraphics[width=0.32\textwidth]{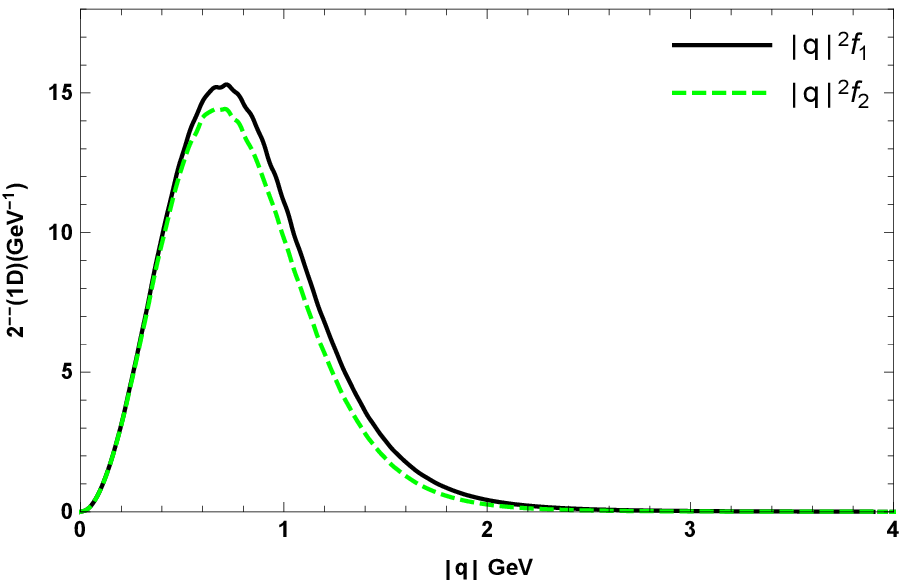}
\includegraphics[width=0.32\textwidth]{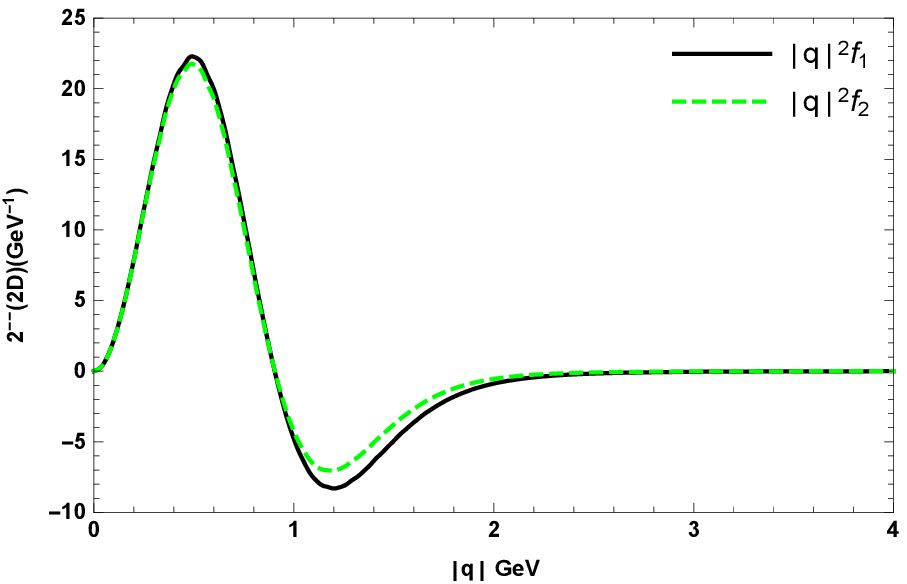}
\includegraphics[width=0.32\textwidth]{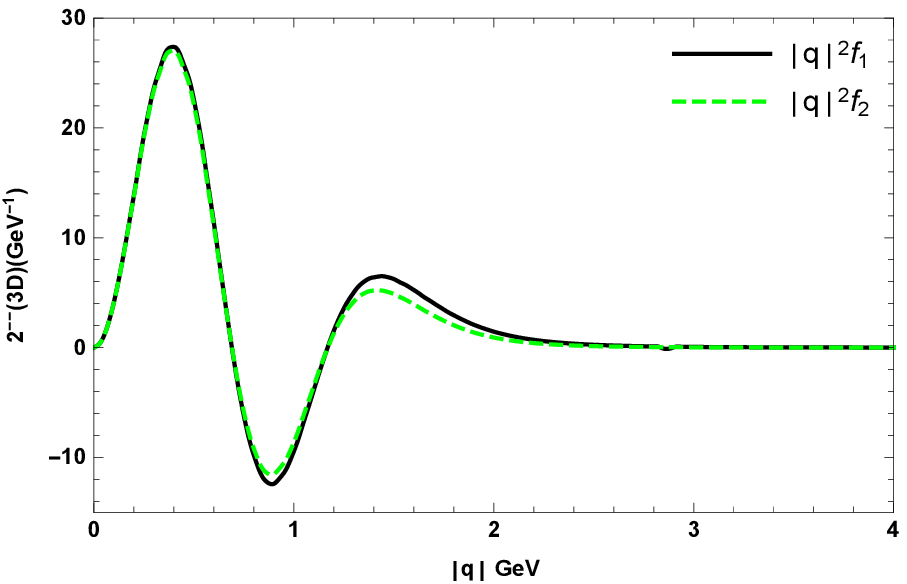}
\caption{{The radial wave functions of the ground, first and second excited $2^{--}$ state $\psi_{_2}(1^{3}D_{_2})$, $\psi_{_2}(2^{3}D_{_2})$ and $\psi_{_2}(3^{3}D_{_2})$.}}\label{fig2-}
\end{figure}

We consider $X(3823)$ as the $2^{--}$ ground state $\psi_{_2}(1^{3}D_{_2})$. From the Eq.(\ref{2--}), it can be seen that, there are two independent radial wave functions $f_{_1}$ and $f_{_2}$. Our results of $f_{_1}$ and $f_{_2}$ are show in Fig. \ref{fig2-}, where instead of $f_{_1}$ and $f_{_2}$, we show the diagrams of $\vec{q}^2f_{_1}$ and $\vec{q}^2f_{_2}$ since they always appear together. From Fig. \ref{fig2-}, we can see clearly that the solution of the $2^{--}$ state has the property $f_{_1}\simeq{f_{_2}}$, this is correct, since in a non-relativistic limit $f_{_1}={f_{_2}}$.

We also show the numerical results of the radial wave functions for excited states $\psi_{_2}(2^{3}D_{_2})$ and $\psi_{_2}(3^{3}D_{_2})$ in Fig. \ref{fig2-}. In general, from the number of nodes of the wave function, we can tell whether the state is a ground state or an excited one. For example, the radial wave function of the ground state has no node, while that of the first excited state has one node and the second excited state has two nodes, etc..

\begin{figure}
\centering
\includegraphics[width=2.6in]{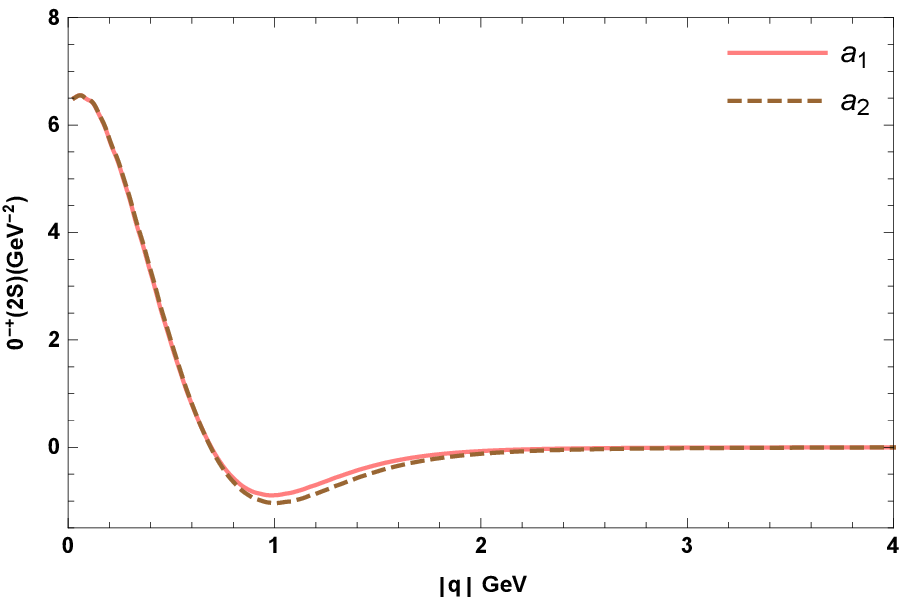}
\includegraphics[width=2.6in]{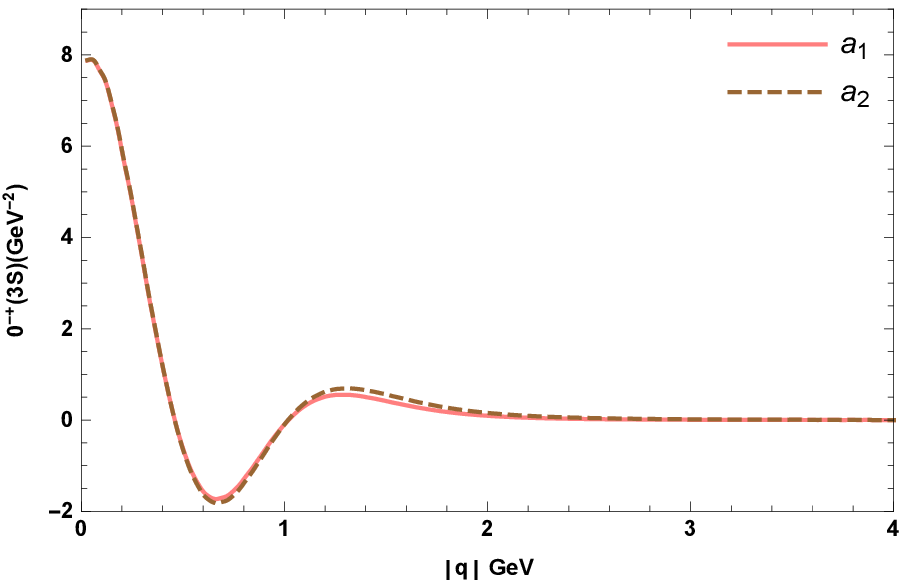}
\caption{The radial wave functions of the $\eta_{_c}(2S)$ and $\eta_{_c}(3S)$.}
\label{f0-}
\end{figure}

\begin{figure}
\centering
\includegraphics[width=2.6in]{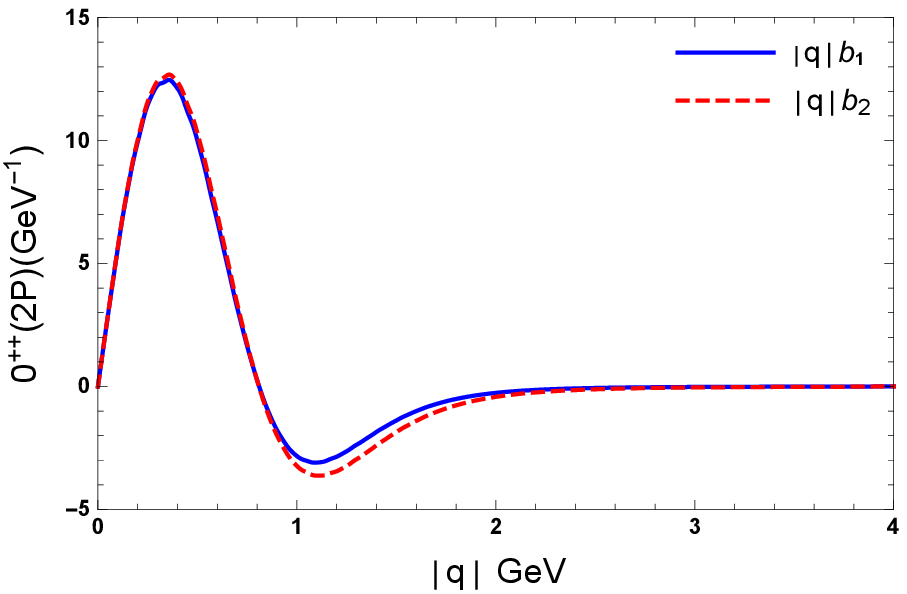}
\includegraphics[width=2.6in]{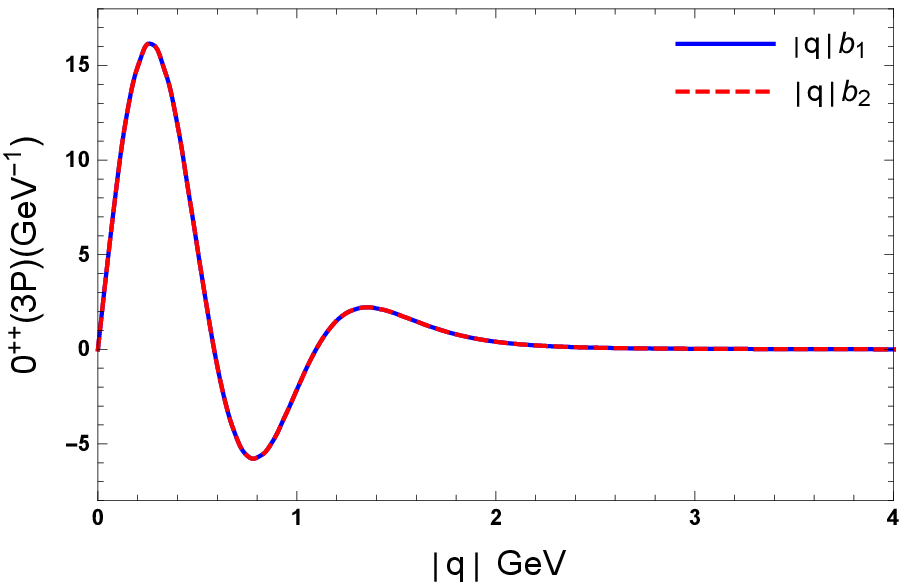}
\caption{{The radial wave functions of the $\chi_{_{c0}}(2P)$ and $\chi_{_{c0}}(3P)$.}}
\label{f0+}
\end{figure}

\begin{figure}
\centering
\includegraphics[width=2.6in]{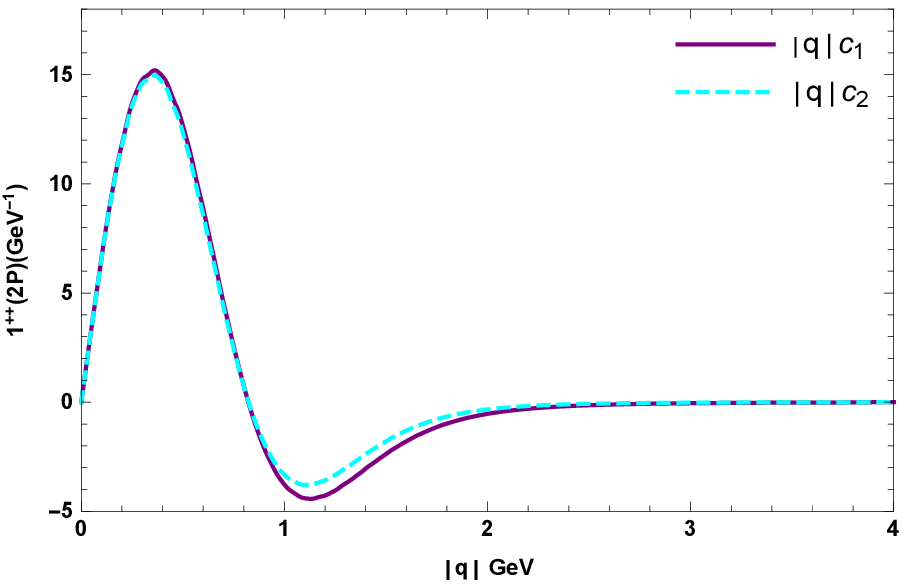}
\includegraphics[width=2.6in]{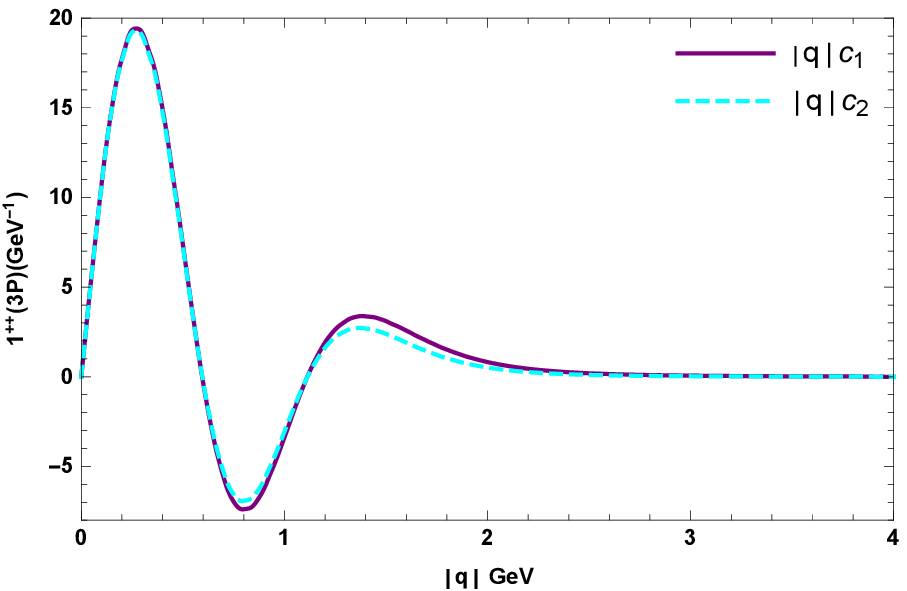}
\caption{The radial wave functions of the $\chi_{_{c1}}(2P)$ and $\chi_{_{c1}}(3P)$.}
\label{f1+}
\end{figure}

\begin{figure}
\centering
\includegraphics[width=2.6in]{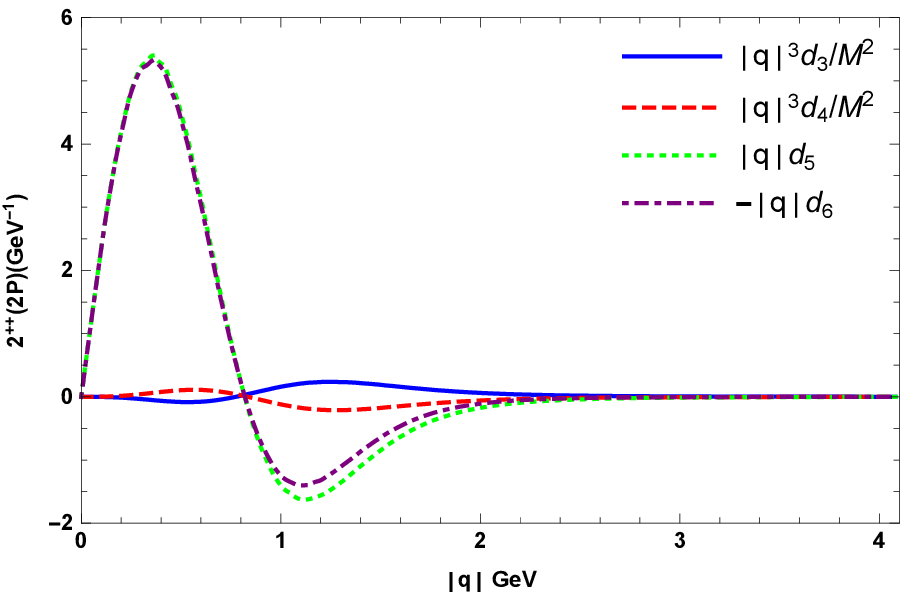}
\includegraphics[width=2.6in]{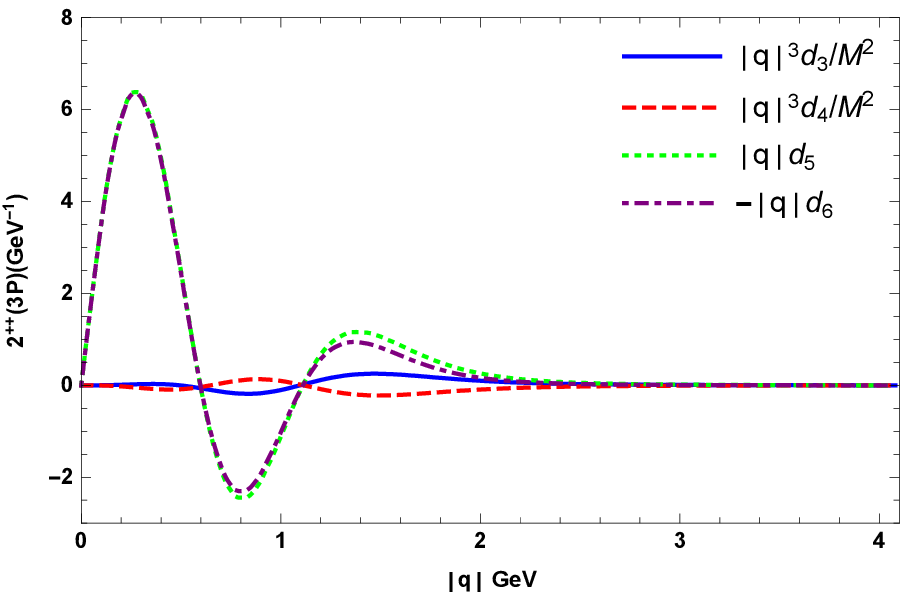}
\caption{The radial wave functions of the $\chi_{_{c2}}(2P)$ and $\chi_{_{c2}}(3P)$.}
\label{f2+}
\end{figure}

\begin{figure}
\centering
\includegraphics[width=2.6in]{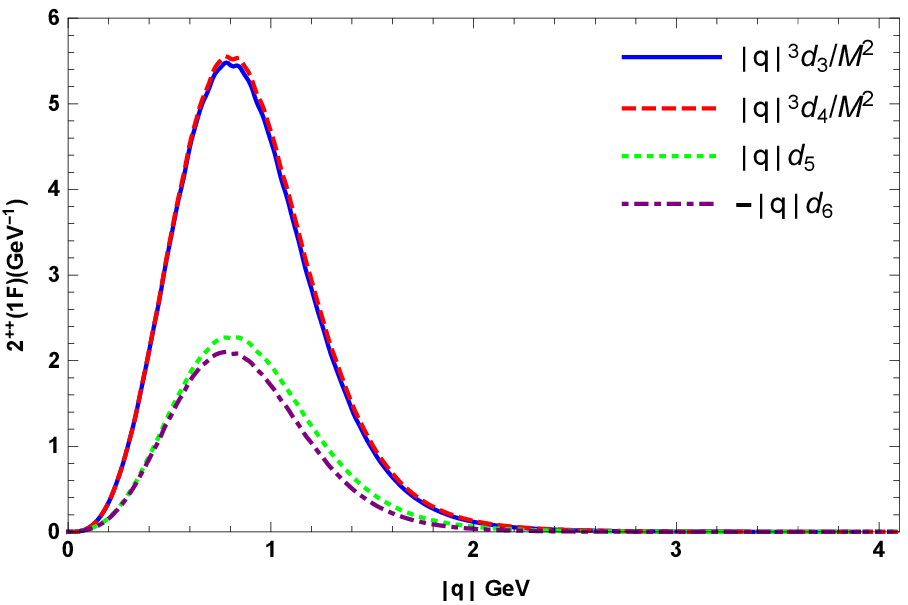}
\includegraphics[width=2.6in]{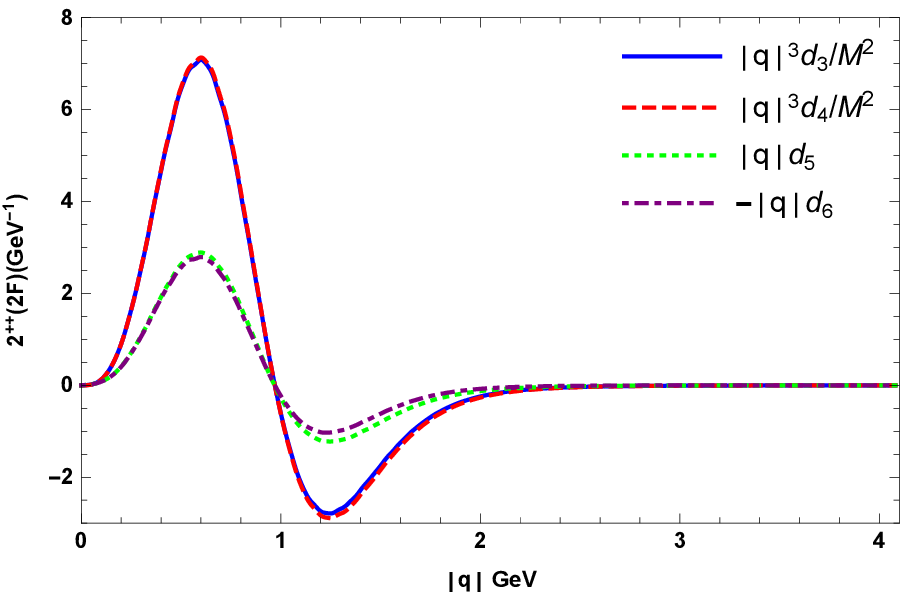}
\caption{The radial wave functions of the $\chi_{_{c2}}(1F)$ and $\chi_{_{c2}}(2F)$.}
\label{f2+F}
\end{figure}

For the $S$ wave states $\eta_{_c}(nS)$ and the $P$ wave states $\chi_{_{c0}}(nP)$, $\chi_{_{c1}}(nP)$ and $\chi_{_{c2}}(nP)$, we have shown the $1S$, $2S$, $1P$ and $2P$ wave functions in previous paper \cite{C-H.Chang.G.L.Wang2010}, but since the theoretical masses of $2^1S_0$ and $2^3P_1$ states are a little different from data, which make the wave functions a little difference from the old ones in Ref. \cite{C-H.Chang.G.L.Wang2010}, we like to show the wave functions for all the excited $S$ and $P$ states one more time in this paper. In Fig. \ref{f0-}, we show the radial wave functions for $\eta_{_c}(2S)$ and $\eta_{_c}(3S)$; in Figs. \ref{f0+}, \ref{f1+} and \ref{f2+}, we give $\chi_{_{cJ}}(2P)$ and $\chi_{_{cJ}}(3P)$, with $J=0,1,2$, respectively; and in Fig. \ref{f2+F}, $\chi_{_{c2}}(1F)$ and $\chi_{_{c2}}(2F)$.

From Figs. \ref{f0-}, \ref{f0+} and \ref{f1+}, we can see that, similar to the $2^{--}$ case, there are two independent radial wave functions for $\eta_{_c}(nS)$, $\chi_{_{c0}}(nP)$ and $\chi_{_{c1}}(nP)$, $n=2,~3$. And they are almost equivalent, this is also confirmed by the non-relativistic limit where they are the same. In Fig. \ref{f2+}, $\chi_{_{c2}}(nP)$ has four independent radial wave functions $d_{_3}$, $d_{_4}$, $d_{_5}$ and $d_{_6}$, where the pure $P$ wave terms $d_{_5}$ and $d_{_6}$ are dominant. And the relation $d_{_5}\simeq -d_{_6}$ is also consistent with the non-relativistic limit $d_{_5}= -d_{_6}$. All other terms are relativistic corrections and they are $D$ and $F$ waves. While in Fig. \ref{f2+F}, for $\chi_{_{c2}}(nF)$, $d_{_3}$ and $d_{_4}(\simeq d_{_3})$ terms are dominant $F$ partial waves, the $P$ waves $d_{_5}$ and $d_{_6}(\simeq -d_{_5})$ terms are sizable, all other terms which are not shown here are $D$ partial waves.

\subsection{EM decay widths of $X(3823)$ as the state $\psi_{_2}(1^{3}D_{_2})$}

{If Eq.(\ref{allamp}), the complete amplitude formula, is used, considering $X(3823)$ as the $\psi_{_2}(1^{3}D_{_2})$ state, the final state are $\eta_{_c}(1S)$ and $\chi_{_{c0}}(1P)$, the decay widths are
\begin{eqnarray}
&&\Gamma[X(3823)\rightarrow\chi_{_{c0}}(1S)\gamma]=1.25 \hspace{0.1cm}\textrm{keV},~~\Gamma[X(3823)\rightarrow\eta_{_c}(1S)\gamma]=1.34 \hspace{0.1cm}\textrm{keV}.
\end{eqnarray}
when Eq.(\ref{amp}) is chosen, that is, only the positive energy wave function contributes to the amplitude, then the decay widths are
\begin{eqnarray}
&&\Gamma[X(3823)\rightarrow\chi_{_{c0}}(1S)\gamma]=1.22 \hspace{0.1cm}\textrm{keV},~~\Gamma[X(3823)\rightarrow\eta_{_c}(1S)\gamma]=1.30 \hspace{0.1cm}\textrm{keV}.
\end{eqnarray}
From the above results, it can be seen that the contributions of the positive energy wave functions $\varphi^{++}$ to the decay width are dominant, and the contributions of other terms are about $2.4\%$ and $3.0\%$ for the two channels. Therefore, in the following calculation, for simplicity, the formula Eq.(\ref{amp}) of decay amplitude is adopted.}

{The EM decay results of other channels for $X(3823)$ ($\psi_{_2}(1^{3}D_{_2})$) are
\begin{eqnarray}
&&\Gamma[X(3823)\rightarrow\chi_{_{\{c1,~c2\}}}(1P)\gamma]=\{265,~~57\} \hspace{0.1cm}\textrm{keV},
\end{eqnarray}
and
\begin{eqnarray}
&&\Gamma[X(3823)\rightarrow\eta_{_c}(2S)\gamma]= 0.069 \hspace{0.1cm}\textrm{keV}.
\end{eqnarray}
We can see the dominant decay channel is $X(3823)\rightarrow\chi_{_{c1}}(1P) \gamma$, and its decay width is much larger than others.}

\begin{table}
\caption{ The decay widths (keV) of the radiative transition $X(3823)\rightarrow\chi_{_{cJ}}(1P)\gamma$ ($J=0,1,2$), $X(3823)\rightarrow\eta_{_c}(1S,~2S)\gamma$ and the ratio of $\frac{\Gamma(\psi_{_2}(1D)\to\chi_{_{c2}}(1P)\gamma)}
{\Gamma(\psi_{_2}(1D)\to\chi_{_{c1}}(1P)\gamma)}$.}
\begin{scriptsize}
%\begin{tiny}
\begin{tabular}{|c|c|c|c|c|c|c|c|c|}
\hline
  & {\cite{C.F.Qiao1997}} & {\cite{E.J.E2002}} & {\cite{D.E2003}} & {\cite{T.B2005}} & {\cite{B.Q.Li2009}} & {\cite{W.J.Deng 2015}} & ours & EX\cite{M. Ablikim2021} \\
  & $RE$ & $RE$ & $NR$  $RV$  $RS$  $RVS$ & $NR$  $GI$ & $NR_1$ $NR_2$  $RE$ & $NR_1$  $NR_2$ & $RE$&\\
\hline
 $\Gamma(\psi_{_2}(1D)\to\chi_{_{c1}}(1P)\gamma)$ & $250$  & $260$ & $297$  $215$ $215$ $215$ & $307$  $268$ & $307$  $342$  $208$ & $285$  $296$ & $265$&\\
 \hline
 $\Gamma(\psi_{_2}(1D)\to\chi_{{c2}}(1P)\gamma)$ & $60$ & $56$ & $62$ ~ $55$ ~ $51$~ $59$ & $64$ ~ $66$ & $64$~ $70$ ~ $55$& $91$~ $96$ & $57$&\\
\hline
$\frac{\Gamma(\psi_{_2}(1D)\to\chi_{_{c2}}(1P)\gamma)}
{\Gamma(\psi_{_2}(1D)\to\chi_{_{c1}}(1P)\gamma)}$(\%) & 24 & 22 & $21$~~$26$~~$24$~~$27$& $21$~~$25$ &$21$~~$20$~~$26$ & $32$~~$32$ &22 &$28^{+14}_{-11}\pm2$\\
\hline
${\Gamma(\psi_{_2}(1D)\to\chi_{_{c0}}(1P)\gamma)}$ &  &  & & & &  & 1.2 &\\
\hline
 $\Gamma(\psi_{_2}(1D)\to\eta_{_c}(1S)\gamma)$ &  &  &  &  &  &  & $1.3$&\\
 \hline
  $\Gamma(\psi_{_2}(1D)\to\eta_{_c}(2S)\gamma)$ &  &  &  &  &  &  & ${0.069(0.067)}$&\\
 \hline
\end{tabular}
\end{scriptsize}
\label{I}
\end{table}

For comparison, we show our results and other model predictions \cite{C.F.Qiao1997,E.J.E2002,D.E2003,T.B2005,B.Q.Li2009,W.J.Deng 2015} in Table \ref{I}. Where, $RE$ represents a relativistic method, $NR$ the non-relativistic method, $GI$ is the relativistic Godfrey-Isgur model, $RV$ and $RS$ represent the relativistic method using vector and scalar potential, respectively, while $RVS$ the mixture of them. {In our results, the value in parentheses is calculated using the experimental mass. It can be seen that the decay width is insensitive to the mass of particle.} We can also see that our results of $X(3823)\rightarrow\chi_{_{\{c1,~c2\}}}(1P)\gamma$ are close to those of relativistic method $RE$ in Refs. \cite{C.F.Qiao1997,E.J.E2002} and relativistic $GI$ model in Ref. \cite{T.B2005}.

In Table \ref{I}, we also show the ratio of the decay rate $X(3823)\rightarrow\chi_{_{c2}}\gamma$ to that of $X(3823)\rightarrow\chi_{_{c1}}\gamma$, our result is
\begin{eqnarray}
&&\frac{{\cal B}[X(3823)\rightarrow\chi_{_{c2}}\gamma]}{{\cal B}[X(3823)\rightarrow\chi_{_{c1}}\gamma]}=22~\%.
\end{eqnarray}
This result and all other theoretical predictions in Table \ref{I} are within the range of current experimental value $0.28^{+0.14}_{-0.11}\pm0.02$ \cite{M. Ablikim2021}. The consistence shows that this ratio cancels some model dependent uncertainties, and it is more reflective of the true value. So we also give the ratio{
\begin{equation}\frac{{\cal B}[X(3823)\rightarrow\chi_{_{c0}}\gamma]}{{\cal B}[X(3823)\rightarrow\chi_{_{c1}}\gamma]} = 0.46~\%.
\end{equation}}
The result is within the experimental limit $<0.24$ detected by BESIII \cite{M. Ablikim2021}. {The channel of $X(3823)\rightarrow\chi_{_{c0}}\gamma$ was also calculated in Ref \cite{W.J.Deng2017}, and they gave a decay width of $1.42$ keV, witch is a little bigger than ours, while their ratio $\frac{{\cal B}[X(3823)\rightarrow\chi_{_{c0}}\gamma]}{{\cal B}[X(3823)\rightarrow\chi_{_{c1}}\gamma]} =0.62~\%$ closes to ours.}

\subsection{EM decay widths of $\psi_{_2}(2^{3}D_{_2})$}

Our predictions for the EM decay widths of the excited state $\psi_{_2}(2D)$ and other theoretical results  are shown in Table \ref{II}. The dominant decay channel is $\psi_{_2}(2D)\rightarrow\chi_{_{c1}}(2P)\gamma$,
\begin{equation}
{\Gamma[\psi_{_2}(2D)\rightarrow\chi_{_{c1}}(2P)\gamma]=237~ \rm{keV}},
\end{equation}
which is close to those of the relativistic $GI$ model in Ref. \cite{T.B2005} and non-relativistic potential model $NR_{_1}$ in Ref. \cite{W.J.Deng 2015}.

\begin{table}
\caption{ The decay widths (keV) of the radiative transition of the $\psi_{_2}(2D)\rightarrow\chi_{_{cJ}}\gamma$ ($J=0,1,2$) and $\psi_{_2}(2D)\rightarrow\eta_{_c}\gamma$. }
\begin{tabular}{|c|c|c|c|c|}
\hline
  & {\cite{T.B2005}} & {\cite{W.J.Deng2017}} & {\cite{W.J.Deng 2015}} & ours  \\
  & $NR$ \hspace{0.5cm} $GI$ & $NR_{1}$ \hspace{0.3cm} $NR_{2}$ \hspace{0.3cm} $NR_{3}$ & $NR_{1}$ \hspace{0.5cm} $NR_{2}$ & $RE$\\
\hline
 $\Gamma(\psi_{_2}(2D)\to\chi_{_{c0}}(1P)\gamma)$ &  \hspace{0.5cm}  &  \hspace{0.8cm}  \hspace{0.8cm}  &  \hspace{0.5cm}  & ${0.16}$\\
 \hline
 $\Gamma(\psi_{_2}(2D)\to\chi_{_{c1}}(1P)\gamma)$ & $26$ \hspace{0.5cm} $23$ & $17$ \hspace{0.8cm} $26$ \hspace{0.8cm} $10$ & $68$ \hspace{0.5cm} $68$ & $33$\\
 \hline
 $\Gamma(\psi_{_2}(2D)\to\chi_{_{c2}}(1P)\gamma)$ & $7.2$ \hspace{0.3cm} $0.62$ & $6.7$ \hspace{0.8cm} $10$ \hspace{0.8cm} $3.8$ & $20$ \hspace{0.5cm} $20$ & $7.3$\\
 \hline
 $\Gamma(\psi_{_2}(2D)\to\chi_{_{c2}}(1F)\gamma)$ &  &  &  & $6.2$\\
 \hline
 $\Gamma(\psi_{_2}(2D)\to\chi_{_{c0}}(2P)\gamma)$ &  \hspace{0.3cm}  &  \hspace{0.6cm}  \hspace{0.6cm}  & \hspace{0.4cm}  & ${1.13}$\\
 \hline
 $\Gamma(\psi_{_2}(2D)\to\chi_{_{c1}}(2P)\gamma)$ & $298$ \hspace{0.3cm} $225$ & $140$ \hspace{0.6cm} $178$ \hspace{0.6cm} $92$ & $223$ \hspace{0.4cm} $188$ & ${237~(230)}$\\
 \hline
 $\Gamma(\psi_{_2}(2D)\to\chi_{_{c2}}(2P)\gamma)$ & $52$ \hspace{0.5cm} $65$ & $39$ \hspace{0.8cm} $64$ \hspace{0.8cm} $19$ & $115$ \hspace{0.5cm} $64$ & $58$\\
\hline
$\frac{\Gamma(\psi_{_2}(2D)\to\chi_{_{c1}}(1P)\gamma)}
{\Gamma(\psi_{_2}(2D)\to\chi_{_{c1}}(2P)\gamma)}$(\%) & $8.7$ \hspace{0.5cm} $10$ & $12$ \hspace{0.8cm} $15$ \hspace{0.8cm} $11$ & $30$ \hspace{0.5cm} $36$ & $14$\\
\hline
$\frac{\Gamma(\psi_{_2}(2D)\to\chi_{_{c2}}(2P)\gamma)}
{\Gamma(\psi_{_2}(2D)\to\chi_{_{c1}}(2P)\gamma)}$(\%) & $17$ \hspace{0.5cm} $29$ & $28$ \hspace{0.8cm} $36$ \hspace{0.8cm} $21$ & $52$ \hspace{0.5cm} $34$ & $25$\\
\hline
$\Gamma(\psi_{_2}(2D)\to\eta_{_c}(1S)\gamma)$ &  \hspace{0.3cm}  &  \hspace{0.6cm}  \hspace{0.6cm}  & \hspace{0.4cm}  & $2.1$\\
\hline
$\Gamma(\psi_{_2}(2D)\to\eta_{_c}(2S)\gamma)$ &  \hspace{0.3cm}  &  \hspace{0.6cm}  \hspace{0.6cm}  & \hspace{0.4cm}  & ${0.33(0.32)}$\\
\hline
$\Gamma(\psi_{_2}(2D)\to\eta_{_c}(3S)\gamma)$ &  \hspace{0.3cm}  &  \hspace{0.6cm}  \hspace{0.6cm}  & \hspace{0.4cm}  & $0.092$\\
\hline
\end{tabular}
\label{II}
\end{table}
{If instead of using the theoretical mass of $\chi_{_{c1}}(2P)$, the experimental value is used, then the decay width for $\psi_{_2}(2D)\to\chi_{_{c1}}(2P)$ becomes to $230$ keV, see the value in parenthesis
in Table \ref{II}. Combined with the result of $\psi_{_2}(2D)\to\eta_{_{c}}(2S)$ in Table \ref{II}, two groups of values are also given, we confirm the previous conclusion that the radiative electromagnetic decay width is not very sensitive to the mass.}

The channels $\psi_{_2}(2D)\to\chi_{_{c1}}(1P)\gamma$ and $\psi_{_2}(2D)\to\chi_{_{c2}}(2P)\gamma$ also have sizable contributions, so we also calculate their decay ratios to the channel $\psi_{_2}(2D)\to\chi_{_{c1}}(2P)\gamma$, and list them in Table \ref{II}. We can see that, unlike the case of $\psi_{_2}(1D)$, the ratios of $\psi_{_2}(2D)$ are much different from model to model. The reason may due to the relativistic corrections being not included or fully considered, because in previous paper \cite{wang2022}, we have pointed out that higher excited states have much larger relativistic corrections than those of lower excited and ground states. This conclusion has been confirmed in the weak transition process \cite{gengzk}.

\subsection{EM decay widths of $\psi_{_2}(3^{3}D_{_2})$}

The predictions for the EM decay width of excited state $\psi_{_2}(3D)$ are shown in Table \ref{III}. The dominant decay channel is $\psi_{_2}(3D)\rightarrow\chi_{_{c1}}(3P)$,
\begin{equation} \Gamma[\psi_{_2}(3D)\rightarrow\chi_{_{c1}}(3P)\gamma]=218~ \rm{keV}.
\end{equation}
We can see that, the dominant EM decay channel for $\psi_{_2}(nD)$ is $\chi_{_{c1}}(nP)\gamma$, and the second is $\chi_{_{c2}}(nP)\gamma$, where $n=1,2,3$, respectively, while $\chi_{_{c0}}(nP)\gamma$ and $\eta_{_{c}}(nS)\gamma$ channels always have small contributions.
\begin{table}
\caption{The EM decay widths (keV) of the excited state $\psi_{_2}(3D)$.}
\begin{tabular}{|c|c|c|c|c|c|c|}
\hline
 Initial state & Final state & $\Gamma_{(our)}$ & Final state & $\Gamma_{(our)}$ & Final state & $\Gamma_{(our)}$ \\
\hline
 $\psi_{_2}(3D)$ & $\chi_{_{c0}}(1P)$ $\gamma$ & ${0.26}$ & $\chi_{_{c0}}(2P)$ $\gamma$ & ${0.54}$ & $\chi_{_{c0}}(3P)$ $\gamma$ & ${1.1}$ \\
\hline
 $\psi_{_2}(3D)$ & $\chi_{_{c1}}(1P)$ $\gamma$ & $38$ & $\chi_{_{c1}}(2P)$ $\gamma$ & ${40(41)}$ & $\chi_{_{c1}}(3P)$ $\gamma$ & $218$ \\
 \hline
 $\psi_{_2}(3D)$ & $\chi_{_{c2}}(1P)$ $\gamma$ & $6.8$ & $\chi_{_{c2}}(2P)$ $\gamma$ & $8.3$ & $\chi_{_{c2}}(3P)$ $\gamma$ & $41$ \\
 \hline
 $\psi_{_2}(3D)$ & $\chi_{_{c2}}(1F)$ $\gamma$ & $8.3$ & $\chi_{_{c2}}(2F)$ $\gamma$ & $11$ &  &  \\
 \hline
$\psi_{_2}(3D)$ & $\eta_{_c}(1S)$ $\gamma$ & $4.6$ & $\eta_{_c}(2S)$ $\gamma$ & ${2.55(2.44)}$ & $\eta_{_c}(3S)$ $\gamma$ & $0.24$ \\
\hline
\end{tabular}
\label{III}
\end{table}

\subsection{Contributions of different partial waves}

In a previous work \cite{wang2022}, we point out that, in a complete relativistic method, the relativistic wave function for a $J^P$ state is not a pure wave. This conclusion is also valid for the charmonium. For the $X(3823)$ as the $2^{--}$ state $\psi_{_2}(1^3D_{_2})$, besides the main $D$ wave, it also includes a small part of $F$ wave; for the $\eta_{_{c}}(1S)$, it is dominated by $S$ wave with a small amount of $P$ partial wave, while for the $\chi_{_{c0}}(1P)$ state, as a $P$ wave dominant state, it includes a small component of $S$ wave, etc, see the details in Sec.II.B.

\begin{table}
\caption{The decay width (keV) of different partial waves for $\psi_{_2}(1D)\to\eta_{_{c}}(1S)\gamma$.}
\begin{tabular}{|c|c|c|c|}
\hline
 \diagbox{$2^{--}$}{$0^{-+}$} & $~complete~$ & $S ~ wave ~ (A_{_{f_1}}, A_{_{f_2}})$ & $P ~ wave ~ (A_{_{f_3}})$ \\
 \hline
 $~complete~$ & $1.3$ & $0.0035$ & $1.3$ \\
\hline
 $~~D ~~ wave ~~ (F_{_{1}},F_{_{2}})~~$ & $3.1$  & $0.41$ & $1.3$ \\
\hline
 $~~F ~~ wave ~~ (F_{_{3}})~~$ & $0.39$ & $0.39$ & $ 0$ \\
 \hline
\end{tabular}
\label{IV}
\end{table}

In this subsection, we study the contributions of different partial waves of the initial and finial mesons to the decay width. The results are shown in Table \ref{IV}$ \sim$ \ref{IX}, where `$complete$' means the complete or whole wave function is used, `$S~wave$' means only the $S$ partial wave has contribution and other partial waves are deleted. From these tables, we can see that in all the decays, the main contribution of $2^{--}$ state $\psi_{_2}$ comes from its dominant partial wave, namely $D$ wave, which is also its non-relativistic term, and its relativistic correction term, namely $F$ partial wave, has a relatively small contribution.

\begin{table}
{\caption{The EM decay width (keV) of different partial waves for $\psi_{_2}(1D)\to\chi_{_{c0}}(1P)\gamma$.}
\begin{tabular}{|c|c|c|c|}\hline
 \diagbox{$2^{--}$}{$0^{++}$} & $~complete~$ & $S ~ wave ~ (B_{_{f_1}})$ & $P ~ wave ~ (B_{_{f_2}}, B_{_{f_3}})$ \\ \hline
 $~complete~$ & $1.2$ & $1.3$ & $0.19$ \\\hline
 $~~D ~~ wave ~~ (F_{_{1}},F_{_{2}})~~$ & $1.3$  & $1.4$ & $0.19$ \\
\hline
 $~~F ~~ wave ~~ (F_{_{3}})~~$ & $0.14$ & $0.14$ & $\sim 0$ \\ \hline
\end{tabular}
\label{V}}
\end{table}

Table \ref{IV} shows the case of $\psi_{_2}(1D)\to\eta_{_{c}}(1S)\gamma$. We know that $\eta_{_{c}}(1S)$ is a $S$-wave dominant state, which only contains a small amount of $P$ partial wave. But from Table \ref{IV}, we can see that the contribution of $D~wave \to S~ wave$ transition is suppressed, indicates that the major contribution of this decay process is due to relativistic effect (dominant by $D~wave \to P~ wave$ transition).

\begin{table}
\caption{The EM decay width (keV) of different partial waves for $\psi_{_2}(1D)\to\chi_{_{c1}}(1P)\gamma$.}
\begin{tabular}{|c|c|c|c|}
\hline
 \diagbox{$2^{--}$}{$1^{++}$} & $~complete~$ & $P ~ wave ~ (C_{_{f_1}}, C_{_{f_2}})$ & $D ~ wave ~ (C_{_{f_3}})$ \\
 \hline
 $~complete~$ & $265$ & $204$ & $4.0$ \\
\hline
 $~~D ~~ wave ~~ (F_{_{1}},F_{_{2}})~~$ & $209$  & $211$ & $4.2$ \\
\hline
 $~~F ~~ wave ~~ (F_{_{3}})~~$ & $3.4$ & $0.17$ & $0.0056$ \\
 \hline
\end{tabular}
\label{VI}
\end{table}

{Table \ref{V} shows the result of $\psi_{_2}(1D)\to\chi_{_{c0}}(1P)\gamma$. This result is similar to the case of $\psi_{_2}(1D)\to\eta_{_{c}}(1S)\gamma$, the contribution of dominant $P$ wave in final state is very small, while the contribution of the small component of $S$ wave is large. From the form factor formula, Eq.(\ref{B2}), we can see the origin of this result. The $P$ wave term of the unique additive relation, $F_{_2}(B_{_{f_3}}+B^{'}_{_{f_3}})(3\cos^{2}\theta-1)$, is suppressed due to the angle integral. The rest have subtractive relationships, $B_{_{f_2}}-B^{'}_{_{f_2}}$ and $B_{_{f_3}}-B^{'}_{_{f_3}}$, therefore their contributions are also suppressed. And in the non-relativistic limit, the contribution of all these $P$ wave terms is zero. So for the EM decay $\psi_{_2}(1D)\to\chi_{_{c0}}(1P)\gamma$, the contribution of $S$ wave which provides the relativistic correction is greater than the that of $P$ wave.}

{Table \ref{VI} show the result of $\psi_{_2}(1D)\to\chi_{_{c1}}(1P)\gamma$. We can see that, the main contribution of the final state come from the dominant $P$ partial wave which provides the non-relativistic result, and the relativistic correction ($D$ partial wave in $1^{++}$ state) contribute very small. The form factors for this decay are shown in Appendix, but they are very complicated, we will not discuss the details.}

Tables \ref{VII}, \ref{VIII} and \ref{IX} show the results of $\psi_{_2}(1D)\to\chi_{_{c2}}(1P)\gamma$, $\psi_{_2}(2D)\to\chi_{_{c2}}(1P)\gamma$ and  $\psi_{_2}(2D)\to\chi_{_{c2}}(1F)\gamma$, respectively, where three final mesons are all $2^{++}$ states. The first two are $1P$ wave dominant states combined with small $D$ and $F$ partial waves, the third one is $1F$ wave dominant state but combined with sizable $P$ and $D$ partial waves \cite{wang2022}. Tables \ref{VII} and \ref{VIII} show us that compared with the dominant $P$ wave, the contributions of $D$ and $F$ partial waves in $1P$ dominant final state are small, and the nodal structure in the wave function of $\psi_{_2}(2D)$ results in the smaller decay width of $\psi_{_2}(2D)\to\chi_{_{c2}}(1P)\gamma$ compared with $\psi_{_2}(1D)\to\chi_{_{c2}}(1P)\gamma$. From Table \ref{IX}, we can see that besides the large contribution of $F$ wave in the $1F$ dominant state, the contribution of $D$ partial wave is also large, but those of $P$ wave are suppressed.

\begin{table}
\caption{The EM decay width (keV) of different partial waves for $\psi_{_2}(1D)\to\chi_{_{c2}}(1P)\gamma$.}
\begin{tabular}{|c|c|c|c|c|}
\hline
 \diagbox{$2^{--}$}{$2^{++}$} & $~complete~$ & $P ~ wave ~ (D_{_{f_5}}, D_{_{f_6}})$ & $D ~ wave ~ (D_{_{f_1}}, D_{_{f_2}}, D_{_{f_7}})$ & $F ~ wave ~ (D_{_{f_3}}, D_{_{f_4}})$ \\
 \hline
 $~complete~$ & $57$ & $18$ & $1.5$ & $0.23$\\
\hline
 $~~D ~~ wave ~~ (F_{_{1}},F_{_{2}})~~$ & $75$  & $44$ & $4.9$ & $0.70$ \\
\hline
 $~~F ~~ wave ~~ (F_{_{3}})~~$ & $1.7$ & $6.1$ & $1.4$  & $0.0057$ \\
 \hline
\end{tabular}
\label{VII}
\end{table}

\begin{table}
\caption{The EM decay width (keV) of different partial waves for $\psi_{_2}(2D)\to\chi_{_{c2}}(1P)\gamma$.}
\begin{tabular}{|c|c|c|c|c|}
\hline
 \diagbox{$2^{--}$}{$2^{++}$} & $~complete~$ & $P ~ wave ~ (D_{_{f_5}}, D_{_{f_6}})$ & $D ~ wave ~ (D_{_{f_1}}, D_{_{f_2}}, D_{_{f_7}})$ & $F ~ wave ~ (D_{_{f_3}}, D_{_{f_4}})$ \\
 \hline
 $~complete~$ & $7.3$ & $3.4$ & $0.39$ & $0.046$\\
\hline
 $~~D ~~ wave ~~ (F_{_{1}},F_{_{2}})~~$ & $9.2$  & $4.9$ & $0.56$ & $0.066$ \\
\hline
 $~~F ~~ wave ~~ (F_{_{3}})~~$ & $0.38$ & $0.24$ & $0.037$  & $0.00028$ \\
 \hline
\end{tabular}
\label{VIII}
\end{table}

\begin{table}
\caption{The EM decay width (keV) of different partial waves for $\psi_{_2}(2D)\to\chi_{_{c2}}(1F)\gamma$.}
\begin{tabular}{|c|c|c|c|c|}
\hline
 \diagbox{$2^{--}$}{$2^{++}$} & $~complete~$ & $P ~ wave ~ (D_{_{f_5}}, D_{_{f_6}})$ & $D ~ wave ~ (D_{_{f_1}}, D_{_{f_2}}, D_{_{f_7}})$ & $F ~ wave ~ (D_{_{f_3}}, D_{_{f_4}})$ \\
 \hline
 $~complete~$ & $6.2$ & $0.65$ & $3.6$ & $5.6$\\
\hline
 $~~D ~~ wave ~~ (F_{_{1}},F_{_{2}})~~$ & $4.8$  & $0.4$ & $2.9$ & $4.3$ \\
\hline
 $~~F ~~ wave ~~ (F_{_{3}})~~$ & $0.55$ & $0.055$ & $0.12$  & $0.46$ \\
 \hline
\end{tabular}
\label{IX}
\end{table}

{If we only keep the dominant partial waves in wave functions and ignore the small partial waves which provide us relativistic corrections for both the initial and final states, then we obtain the non-relativistic results,
\begin{eqnarray}
&&\Gamma_{_0}[X(3823)(1D)\rightarrow\eta_{_{c}}(1S)\gamma]=0.41\hspace{0.1cm}\textrm{keV},
\end{eqnarray}
\begin{eqnarray}
&&\Gamma_{_0}[X(3823)(1D)\rightarrow\chi_{_{\{c0,~c1,~c2\}}}(1P)\gamma]=\{0.19,~211,~44\} \hspace{0.1cm}\textrm{keV},
\end{eqnarray}
Compared with the complete relativistic results, the relativistic effects (defined as $\frac{\Gamma-\Gamma_0}{\Gamma}$) make up 68$~\%$, 84$~\%$, 20$~\%$, 23$~\%$ of $X(3823)\rightarrow\eta_{_{c}}(1S)\gamma$, $X(3823)\rightarrow\chi_{_{cJ}}(1P)\gamma$ ($J=0,1,2$), respectively. So the contribution of the relativistic correction plays a leading role in the decay processes of $\psi_{_2}(1D)\to\eta_{_{c}}(1S)\gamma$ and $\psi_{_2}(1D)\to\chi_{_{c0}}(1P)\gamma$.}
\subsection{Discussion and Conclusion}

In a previous paper \cite{T.H.Wang2016}, we have estimated the annihilation decay (including $ggg$ and $gg\gamma$ final states) width of $X(3823)$, which is about $9.8$ keV. {From Eichten's work \cite{E.J.E2002}, we can get the decay width $\Gamma[\psi_{_2}(^3D_{_2})\to J/\psi\pi\pi]\approx45~$keV. So the total decay width of $X(3823)$ can be estimated as,
\begin{equation}
\Gamma[X(3823)]\approx \Gamma(\eta_{_c}\gamma)+\sum\Gamma(\chi_{_{cJ}}\gamma)+\Gamma(J/\psi\pi\pi)
+\Gamma(ggg)+\Gamma(gg\gamma)\approx 379~\rm{keV}.
\end{equation}}
{Therefore, the process $X(3823)\to\chi_{_{c1}}\gamma$ whose partial width is estimated as $265$ keV, is the dominant decay channel of $X(3823)$. The detection of this channel in experiment is crucial to confirm $X(3823)$ being the state $\psi_{_2}(^3D_{_2})$.}

In conclusion, we study the EM decays of $\psi_{_2}(n^3D_{_2})$ ($n=1,2,3$) by using the relativistic Bethe-Salpeter method, where the new particle $X(3823)$ is treated as $\psi_{_2}(1^3D_{_2})$ in this paper. We find for $\psi_{_2}(n^3D_{_2})$, the dominant EM decay channel is $\psi_{_2}(n^3D_{_2})\to\chi_{_{c1}}(nP)\gamma$. Our results show that $\Gamma[X(3823)\rightarrow\chi_{_{c1}}\gamma]=265$ keV, {compared with the estimated total width $\Gamma[X(3823)]\approx 379$ keV,} this is the dominant decay channel. The decay ratio ${\cal B}[X(3823)\rightarrow\chi_{_{c2}}\gamma]/{\cal B}[X(3823)\rightarrow\chi_{_{c1}}\gamma]=22\%$ is consistent with the observation $0.28^{+0.14}_{-0.11}\pm0.02$, { and the decay ratio ${\cal B}[X(3823)\rightarrow\chi_{_{c0}}\gamma]/{\cal B}[X(3823)\rightarrow\chi_{_{c1}}\gamma]$ $\simeq 0.46\%$ is also less than experimental upper limit $0.24$. In addition, we calculated the contributions of different partial waves. For the decays $X(3823)\to\eta_{_{c}}(1S)\gamma$ and $X(3823)\to\chi_{_{c0}}(1P)\gamma$, the main contribution comes from the relativistic effect, while for the $X(3823) \to \chi_{_{cJ}}(1P)\gamma$ ($J=1, 2$) decay, the non-relativistic contribution is the dominant one. These results may provide useful information to reveal the nature of $X(3823)$ as the Charmonium $\psi_{_2}(1^3D_{_2})$.}

{\bf Acknowledgments}
This work was supported in part by the National Natural Science Foundation of China (NSFC) under the Grants Nos. 12075073, 11865001, 12075074, the Natural Science Foundation of Hebei province under the Grant No. A2021201009, Post-graduate's Innovation Fund Project of Hebei University under the Grant No. HBU2022BS002.

\appendix

\section{The integrals over the relative momentum}
When calculating the integral with respect to $q_{_\perp}$ in the amplitude Eq.(5), we apply the following formula
\begin{eqnarray}\label{2--2}
&&\int\frac{d^3q}{(2\pi)^3}q^{\mu}_{_\perp}F\equiv f_{_{11}}P^{\mu}_{_{f_\perp}},~~~\int\frac{d^3q}{(2\pi)^3}q^{\mu}_{_\perp}q^{\nu}_{_\perp}F\equiv f_{_{21}}P^{\mu}_{_{f_\perp}}P^{\nu}_{_{f_\perp}}+f_{_{22}}g^{\mu\nu}_{_\perp},
\nonumber\\&&\int\frac{d^3q}{(2\pi)^3}q^{\mu}_{_\perp}q^{\nu}_{_\perp}q^{\alpha}_{_\perp}F\equiv f_{_{31}}P^{\mu}_{_{f_\perp}}P^{\nu}_{_{f_\perp}}P^{\alpha}_{_{f_\perp}}+f_{_{32}}(P^{\mu}_{_{f_\perp}}g^{\nu\alpha}_{_\perp}
+P^{\nu}_{_{f_\perp}}g^{\mu\alpha}_{_\perp}+P^{\alpha}_{_{f_\perp}}g^{\mu\nu}_{_\perp}),
\nonumber\\&&\int\frac{d^3q}{(2\pi)^3}q^{\mu}_{_\perp}q^{\nu}_{_\perp}q^{\alpha}_{_\perp}q^{\beta}_{_\perp}F\equiv f_{_{41}}P^{\mu}_{_{f_\perp}}P^{\nu}_{_{f_\perp}}P^{\alpha}_{_{f_\perp}}P^{\beta}_{_{f_\perp}}+f_{_{42}}(P^{\mu}_{_{f_\perp}}P^{\nu}_{_{f_\perp}}g^{\alpha\beta}_{_\perp}+P^{\mu}_{_{f_\perp}}P^{\alpha}_{_{f_\perp}}g^{\nu\beta}_{_\perp}
\nonumber\\&&\hspace{4.5cm}+P^{\mu}_{_{f_\perp}}P^{\beta}_{_{f_\perp}}g^{\nu\alpha}_{_\perp}+P^{\nu}_{_{f_\perp}}P^{\alpha}_{_{f_\perp}}g^{\mu\beta}_{_\perp}+P^{\nu}_{_{f_\perp}}P^{\beta}_{_{f_\perp}}g^{\mu\alpha}_{_\perp}+P^{\alpha}_{_{f_\perp}}P^{\beta}_{_{f_\perp}}g^{\mu\nu}_{_\perp})
\nonumber\\&&\hspace{4.5cm}+f_{_{43}}(g^{\mu\nu}_{_\perp}g^{\alpha\beta}_{_\perp}+g^{\mu\alpha}_{_\perp}g^{\nu\beta}_{_\perp}+g^{\mu\beta}_{_\perp}g^{\nu\alpha}_{_\perp}),
\end{eqnarray}
where $F=F(q^2_{_\perp},q^2_{_{f_\perp}})$, and we have used the following abbreviations
\begin{eqnarray}
&&P^{\mu}_{_{f_\perp}}\equiv P^{\mu}_{_{f}}-\frac{P\cdot P_{_f}}{M^2}P^{\mu},~~~g^{\mu\nu}_{_\perp}\equiv g^{\mu\nu}-\frac{P^{\mu}P^{\nu}}{M^2}.
\end{eqnarray}
The coefficient $f_{_{ij}}$ are calculated as
\begin{eqnarray}
&&f_{_{11}}=\int\frac{d^3q}{(2\pi)^3}F\frac{q}{r}\cos\theta,
~~f_{_{21}}=\int\frac{d^3q}{(2\pi)^3}F\frac{q^2}{2r^2}(3\cos^{2}\theta-1),
\nonumber\\&&f_{_{22}}=\int\frac{d^3q}{(2\pi)^3}F\frac{q^2}{2}(\cos^{2}\theta-1),
~~f_{_{31}}=\int\frac{d^3q}{(2\pi)^3}F\frac{q^3}{2r^3}(5\cos^{3}\theta-3\cos\theta),
\nonumber\\&&f_{_{32}}=\int\frac{d^3q}{(2\pi)^3}F\frac{q^3}{2r}(\cos^{3}\theta-\cos\theta),
~~f_{_{41}}=\int\frac{d^3q}{(2\pi)^3}F\frac{q^4}{8r^4}(35\cos^{4}\theta-30\cos^{2}\theta+3),
\nonumber\\&&f_{_{42}}=\int\frac{d^3q}{(2\pi)^3}F\frac{q^4}{8r^2}(5\cos^{4}\theta-6\cos^{2}\theta+1),
\nonumber\\&&f_{_{43}}=\int\frac{d^3q}{(2\pi)^3}F\frac{q^4}{8}(\cos^{4}\theta-2\cos^{2}\theta+1),
\end{eqnarray}
where $\theta$ is the angle between $\vec{ q}$ and $\vec{ P_{_f}}$, we have defined $q\equiv\vec{\mid q\mid}$ and $r\equiv\vec{\mid P_{_f}\mid}$.
\section{Form factors}

Here, we will give the detailed expression of the form factors in the corresponding decay channel. For the decay channel $X(3823)$ $\rightarrow$ $\eta_{_c}(^1S_{_0})\gamma$, the form factors $h{_{_1}}$ and $h{_{_2}}$ are
\begin{eqnarray}\label{B1}
&&h_{_1}=\int\frac{q^2~d{q}~d{\cos\theta}}{(2\pi)^{2}}4\bigg\{\frac{q^2}{MM_{_f}}\bigg[\frac{F_{_3}}{M}(A_{_{f_2}}-A^{'}_{_{f_2}})\bigg]\frac{q}{|\vec{P_{_f}}|}\cos\theta+\frac{1}{MM_{_f}}\bigg[\frac{F_{_2}}{M_{_f}}(A_{_{f_3}}-A^{'}_{_{f_3}})
\nonumber\\&&\hspace{1cm}+\frac{F_{_3}}{M}(A_{_{f_2}}-A^{'}_{_{f_2}})\bigg]\frac{q^{3}}{2|\vec{P_{_f}}|}(5\cos^{3}\theta-3\cos\theta)-\frac{2}{MM_{_f}}\bigg[\frac{F_{_3}}{M}(A_{_{f_2}}-A^{'}_{_{f_2}})\bigg]\frac{q^{3}}{2|\vec{P_{_f}}|}\times
\nonumber\\&&\hspace{1cm}(\cos^{3}\theta-\cos\theta)+\frac{1}{M_{_f}}\bigg[F_{_1}(A_{_{f_2}}+A^{'}_{_{f_2}})+\frac{F_{_2}\alpha_{_{f}}E_{_f}}{M_{_f}}(A_{_{f_3}}+A^{'}_{_{f_3}})\bigg](1-\frac{E_{_f}}{M})\frac{q^{2}}{2|\vec{P_{_f}}|^{2}}\times
\nonumber\\&&\hspace{1cm}(3\cos^{2}\theta-1)+\frac{E_{_f}}{M_{_f}}\bigg[\frac{F_{_2}}{M_{_f}}(A_{_{f_3}}-A^{'}_{_{f_3}})+\frac{F_{_3}}{M}(A_{_{f_2}}-A^{'}_{_{f_2}})\bigg](1-\frac{E_{_f}}{M})\frac{q^{3}}{2|\vec{P_{_f}}|^{3}}\times
\nonumber\\&&\hspace{1cm}(5\cos^{3}\theta-3\cos\theta)\bigg\},
\nonumber\\&&h_{_2}=(M^2-ME_{_f})h_{_1},
\end{eqnarray}
where $\alpha_{_{f}}=\alpha_{_{1}}=\alpha_{_{2}}=0.5$.

For the decay channel $X(3823)$ $\rightarrow$ $\chi_{_{c0}}(^3P_{_0})\gamma$, the form factor $t{_{_1}}$ is
\begin{eqnarray}\label{B2}
&&t_{_1}=\int\frac{q^2~d{q}~d{\cos\theta}}{(2\pi)^{2}}4\bigg\{\bigg[\frac{F_{_3}q^2}{M^2M_{_f}}(B_{_{f_1}}+B^{'}_{_{f_1}})+\frac{F_{_2}}{M}(B_{_{f_1}}+B^{'}_{_{f_1}})\bigg]\frac{q^{2}}{2|\vec{P_{_f}}|^2}(3\cos^{2}\theta-1)+
\nonumber\\&&\hspace{1cm}\frac{1}{MM_{_f}}\bigg[\frac{F_{_1}}{M}(B_{_{f_2}}-B^{'}_{_{f_2}})-\frac{F_{_2}}{M_{_f}}(B_{_{f_3}}-B^{'}_{_{f_3}})\bigg]\frac{q^{3}}{2|\vec{P_{_f}}|}(\cos^{3}\theta-\cos\theta)-\frac{1}{MM_{_f}}\bigg[\frac{F_{_3}\alpha_{_f}}{M}\times
\nonumber\\&&\hspace{1cm}(B_{_{f_2}}-B^{'}_{_{f_2}})\bigg]\frac{q^{3}}{2|\vec{P_{_f}}|}(3\cos^{3}\theta-\cos\theta)\bigg\}.
\end{eqnarray}

{\color{red}For the decay channel $X(3823)$ $\rightarrow$ $\chi_{_{c1}}(^3P_{_1})\gamma$, the form factors $s{_{_i}}$ are
\begin{eqnarray}
&&s_{_1}=\int\frac{q^2~d{q}~d{\cos\theta}}{(2\pi)^{2}}4\bigg\{\bigg[\frac{\alpha_{_{f}}F_{_1}E_{_f}}{MM^3_{_f}}\bigg(\alpha_{_{f}}E^2_{_f}(C_{_{f_3}}+C^{'}_{_{f_3}})
-P_{_f}q(C_{_{f_3}}-C^{'}_{_{f_3}})\bigg)+\frac{F_{_2}}{MM^2_{_f}}\bigg(-
\nonumber\\&&\hspace{1cm}\alpha_{_{f}}E^2_{_f}(C_{_{f_2}}+C^{'}_{_{f_2}})+P_{_f}q(C_{_{f_2}}-C^{'}_{_{f_2}})\bigg)-
\frac{F_{_3}q^2}{M^2M_{_f}}(C_{_{f_1}}+C^{'}_{_{f_1}})\bigg]\frac{{q}^{2}}{2|\vec{P_{_f}}|^{2}}(3\cos^{2}\theta-1)
\nonumber\\&&\hspace{1cm}-\bigg[\frac{F_{_1}}{M_{_f}}(C_{_{f_1}}-C^{'}_{_{f_1}})-\frac{F_{_3}\alpha_{_{f}}E_{_f}}{MM_{_f}}(C_{_{f_1}}-C^{'}_{_{f_1}})\bigg]
\frac{E_{_f}}{M}\frac{q^{3}}{2|\vec{P_{_f}}|^{3}}(5\cos^{3}\theta-3\cos\theta)+
\nonumber\\&&\hspace{1cm}\bigg[-\frac{F_{_2}}{M}(C_{_{f_2}}-C^{'}_{_{f_2}})+\frac{F_{_3}}{M^2M_{_f}}\bigg(P_{_f}q(C_{_{f_1}}+C^{'}_{_{f_1}})+
\frac{\alpha_{_{f}}E^2_{_f}P_{_f}q}{M^2_{_f}}(C_{_{f_3}}+C^{'}_{_{f_3}})+q^2\times
\nonumber\\&&\hspace{1cm}(C_{_{f_3}}-C^{'}_{_{f_3}})\bigg)\bigg]\frac{q^{3}}{2|\vec{P_{_f}}|^{3}}(5\cos^{3}\theta-3\cos\theta)+
\bigg[\frac{F_{_3}\alpha_{_{f}}E_{_f}P_{_f}q}{M^3M_{_f}}\bigg(-(C_{_{f_1}}-C^{'}_{_{f_1}})+
\nonumber\\&&\hspace{1cm}\frac{\alpha_{_{f}}E^2_{_f}}{M^2_{_f}}(C_{_{f_3}}-C^{'}_{_{f_3}})\bigg)\bigg]\frac{q^{2}}{2|\vec{P_{_f}}|^{2}}(3\cos^{2}\theta-1)
+\bigg[\frac{F_{_1}}{M_{_f}}\bigg((C_{_{f_1}}-C^{'}_{_{f_1}})+\frac{\alpha_{_{f}}E^2_{_f}}{M^2_{_f}}(C_{_{f_3}}-
\nonumber\\&&\hspace{1cm}C^{'}_{_{f_3}})\bigg)\bigg]\frac{1}{M^2}\frac{q^{3}}{2|\vec{P_{_f}}|}(\cos^{3}\theta-\cos\theta)
+\bigg[\frac{F_{_3}E_{_f}}{MM_{_f}}\bigg((C_{_{f_1}}+C^{'}_{_{f_1}})+\frac{P_{_f}q}{M^2_{_f}}(C_{_{f_3}}-C^{'}_{_{f_3}})\bigg)+
\nonumber\\&&\hspace{1cm}\frac{F_{_1}}{M_{_f}}(C_{_{f_3}}+C^{'}_{_{f_3}})\bigg]\frac{1}{M^2}\frac{q^{4}}{8|\vec{P_{_f}}|^{2}}(5\cos^{4}\theta-6\cos^{2}\theta+1)
+\bigg[\frac{F_{_3}\alpha_{_{f}}E_{_f}q^2}{M^3M_{_f}}\bigg((C_{_{f_1}}-C^{'}_{_{f_1}})-
\nonumber\\&&\hspace{1cm}\frac{\alpha_{_{f}}E^2_{_f}}{M^2_{_f}}(C_{_{f_3}}-C^{'}_{_{f_3}})\bigg)\bigg]\frac{q}{|\vec{P_{_f}}|}\cos\theta
-\bigg[\frac{F_{_1}}{M_{_f}}\bigg(-(C_{_{f_1}}-C^{'}_{_{f_1}})+\frac{\alpha_{_{f}}E^2_{_f}}{M^2_{_f}}(C_{_{f_3}}-C^{'}_{_{f_3}})\bigg)\bigg]\times
\nonumber\\&&\hspace{1cm}\frac{E_{_f}}{M}\frac{q^{3}}{2|\vec{P_{_f}}|^{3}}(5\cos^{3}\theta-3\cos\theta)
-\bigg[\frac{F_{_1}}{MM^3_{_f}}\bigg(\alpha^{2}_{_{f}}E^3_{_f}(C_{_{f_3}}+C^{'}_{_{f_3}})-\alpha_{_{f}}E_{_f}P_{_f}q(C_{_{f_3}}-C^{'}_{_{f_3}})\bigg)
\nonumber\\&&\hspace{1cm}+\frac{F_{_2}}{MM^2_{_f}}\bigg(-\alpha_{_{f}}E^2_{_f}(C_{_{f_2}}+C^{'}_{_{f_2}})+
P_{_f}q(C_{_{f_2}}-C^{'}_{_{f_2}})\bigg)-\frac{F_{_3}q^2}{M^2M_{_f}}(C_{_{f_1}}+C^{'}_{_{f_1}})\bigg]\frac{E_{_f}}{M}\times
\nonumber\\&&\hspace{1cm}\frac{q^{2}}{2|\vec{P_{_f}}|^{2}}(3\cos^{2}\theta-1)
-\bigg[\frac{F_{_3}}{M^2M_{_f}}\bigg(-q^2(C_{_{f_3}}-C^{'}_{_{f_3}})+\alpha^2_{_{f}}E^2_{_f}(C_{_{f_3}}-C^{'}_{_{f_3}})\bigg)+\frac{F_{_2}}{M}\times
\nonumber\\&&\hspace{1cm}(C_{_{f_2}}-C^{'}_{_{f_2}})\bigg]\frac{E_{_f}}{M}\frac{q^{3}}{2|\vec{P_{_f}}|^{3}}(5\cos^{3}\theta-3\cos\theta)
-\bigg[\frac{F_{_3}}{M^2M_{_f}}\bigg(P_{_f}q(C_{_{f_1}}+C^{'}_{_{f_1}})+\frac{\alpha_{_{f}}E^2_{_f}P_{_f}q}{M^2_{_f}}\times
\nonumber\\&&\hspace{1cm}(C_{_{f_3}}+C^{'}_{_{f_3}})+q^2(C_{_{f_3}}-C^{'}_{_{f_3}})\bigg)-\frac{F_{_2}}{M}(C_{_{f_2}}-C^{'}_{_{f_2}})\bigg]
\frac{E_{_f}}{M}\frac{q^{3}}{2|\vec{P_{_f}}|^{3}}(5\cos^{3}\theta-3\cos\theta)+
\nonumber\\&&\hspace{1cm}\bigg[\frac{F_{_1}}{M_{_f}}\bigg(-(C_{_{f_1}}-C^{'}_{_{f_1}})+\frac{\alpha_{_{f}}E^2_{_f}}{M^2_{_f}}(C_{_{f_3}}-C^{'}_{_{f_3}})\bigg)\bigg]
\frac{E^2_{_f}}{M^2}\frac{q^{3}}{2|\vec{P_{_f}}|^{3}}(5\cos^{3}\theta-3\cos\theta)-\bigg[\frac{F_{_1}}{MM_{_f}}
\nonumber\\&&\hspace{1cm}\alpha_{_{f}}E_{_f}(C_{_{f_1}}+C^{'}_{_{f_1}})
+\frac{F_{_2}\alpha_{_{f}}E^2_{_f}}{MM^2_{_f}}(C_{_{f_3}}+C^{'}_{_{f_3}})+\frac{F_{_3}q^2}{M^2M^3_{_f}}P_{_f}q(C_{_{f_3}}-C^{'}_{_{f_3}})\bigg]\frac{E_{_f}}{M}\frac{q^{2}}{2|\vec{P_{_f}}|^{2}}\times
\nonumber\\&&\hspace{1cm}(3\cos^{2}\theta-1)\bigg\},
\end{eqnarray}
where we have defined $P_{_f}q=P_{_f}\cdot q_{_\perp}=-|\vec{P_{_f}}||\vec{q}|\cos\theta$.
\begin{eqnarray}
&&s_{_3}=\int\frac{q^2~d{q}~d{\cos\theta}}{(2\pi)^{2}}4\bigg\{\bigg[\frac{F_{_1}\alpha_{_{f}}E_{_f}}{MM_{_f}}(C_{_{f_1}}+C^{'}_{_{f_1}})
+\frac{F_{_2}\alpha_{_{f}}E^2_{_f}}{MM^2_{_f}}(C_{_{f_3}}+C^{'}_{_{f_3}})+\frac{F_{_3}q^2}{M^2M^3_{_f}}P_{_f}q(C_{_{f_3}}-
\nonumber\\&&\hspace{1cm}C^{'}_{_{f_3}})\bigg]\frac{q^{2}}{2}(\cos^{2}\theta-1)+\bigg[\frac{F_{_1}q^2}{M_{_f}}\bigg(-(C_{_{f_1}}-C^{'}_{_{f_1}})+\frac{P_{_f}q}{M^2_{_f}}(C_{_{f_3}}
+C^{'}_{_{f_3}})\bigg)+\frac{F_{_3}\alpha_{_{f}}E_{_f}q^2}{MM_{_f}}\times
\nonumber\\&&\hspace{1cm}(C_{_{f_1}}-C^{'}_{_{f_1}})\bigg] \frac{q}{|\vec{P_{_f}}|}\cos\theta+\bigg[-\frac{F_{_1}\alpha_{_{f}}E_{_f}P_{_f}q}{MM_{_f}}(C_{_{f_1}}+C^{'}_{_{f_1}})
+\frac{F_{_2}}{M}(\frac{(P_{_f}q)^2}{M^2_{_f}}-q^2)(C_{_{f_2}}-
\nonumber\\&&\hspace{1cm}C^{'}_{_{f_2}})+\frac{F_{_3}q^2}{M^2M_{_f}}\bigg(q^2(C_{_{f_3}}-C^{'}_{_{f_3}})+
\frac{\alpha_{_{f}}E^2_{_f}P_{_f}q}{M^2_{_f}}(C_{_{f_3}}+C^{'}_{_{f_3}})\bigg)\bigg]\frac{q}{|\vec{P_{_f}}|}\cos\theta-\bigg[\frac{F_{_1}}{M_{_f}}\times
\nonumber\\&&\hspace{1cm}\bigg(-\frac{\alpha^2_{_{f}}|\vec{P_{_f}}|^2E^{2}_{_f}}{M^2_{_f}}(C_{_{f_3}}+C^{'}_{_{f_3}})+\frac{P_{_f}q}{M^2_{_f}}
\alpha_{_{f}}E^2_{_f}(C_{_{f_3}}-C^{'}_{_{f_3}})+P_{_f}q(C_{_{f_1}}-C^{'}_{_{f_1}})+q^2\times
\nonumber\\&&\hspace{1cm}(C_{_{f_1}}+C^{'}_{_{f_1}})\bigg)+F_{_2}E_{_f}\bigg(\frac{\alpha_{_{f}}|\vec{P_{_f}}|^2}{M^2_{_f}}
(C_{_{f_2}}+C^{'}_{_{f_2}})-\frac{\alpha_{_{f}}P_{_f}q}{M^2_{_f}}(C_{_{f_2}}-C^{'}_{_{f_2}})\bigg)+
\nonumber\\&&\hspace{1cm}\frac{F_{_3}E_{_f}q^2}{MM_{_f}}(C_{_{f_1}}+C^{'}_{_{f_1}})\bigg]\frac{E_{_f}}{M}\frac{q^{2}}{2|\vec{P_{_f}}|^2}(3\cos^{2}\theta-1)
+\bigg[\frac{F_{_1}}{M_{_f}}\bigg(-(C_{_{f_1}}-C^{'}_{_{f_1}})+
\nonumber\\&&\hspace{1cm}\frac{\alpha_{_{f}}E^2_{_f}}{M^2_{_f}}(C_{_{f_2}}-C^{'}_{_{f_2}})\bigg)\bigg]\frac{q^{3}}{2|\vec{P_{_f}}|}(\cos^{3}\theta-\cos\theta)
-2\bigg[\frac{F_{_3}}{M^2M_{_f}}\bigg(-q^2(C_{_{f_3}}-C^{'}_{_{f_3}})+
\nonumber\\&&\hspace{1cm}\alpha^2_{_{f}}E^2_{_f}(C_{_{f_2}}-C^{'}_{_{f_2}})\bigg)+\frac{F_{_2}}{M}(C_{_{f_2}}-C^{'}_{_{f_2}})\bigg]\frac{q^{3}}{2|\vec{P_{_f}}|}(\cos^{3}\theta-\cos\theta)
-\bigg[\frac{F_{_1}}{M_{_f}}\bigg(-(C_{_{f_1}}-
\nonumber\\&&\hspace{1cm}C^{'}_{_{f_1}})+\frac{\alpha_{_{f}}E^2_{_f}}{M^2_{_f}}(C_{_{f_2}}-C^{'}_{_{f_2}})\bigg)\bigg]\frac{E_{_f}}{M}\frac{q^{3}}{2|\vec{P_{_f}}|}(\cos^{3}\theta-\cos\theta)
+\bigg[\frac{F_{_1}}{M_{_f}}\bigg(-\frac{\alpha^2_{_{f}}|\vec{P_{_f}}|^2E^{2}_{_f}}{M^2_{_f}}\times
\nonumber\\&&\hspace{1cm}(C_{_{f_3}}+C^{'}_{_{f_3}})+\frac{\alpha_{_{f}}E^2_{_f}P_{_f}q}{M^2_{_f}}(C_{_{f_3}}-C^{'}_{_{f_3}})+P_{_f}q(C_{_{f_1}}-C^{'}_{_{f_1}})+q^2(C_{_{f_1}}+C^{'}_{_{f_1}})\bigg)+
\nonumber\\&&\hspace{1cm}F_{_2}E_{_f}\bigg(\frac{\alpha_{_{f}}|\vec{P_{_f}}|^2}{M^2_{_f}}(C_{_{f_2}}+C^{'}_{_{f_2}})-\frac{\alpha_{_{f}}P_{_f}q}{M^2_{_f}}(C_{_{f_2}}-C^{'}_{_{f_2}})\bigg)
+\frac{F_{_3}E_{_f}q^2}{MM_{_f}}(C_{_{f_1}}+C^{'}_{_{f_1}})\bigg]\times
\nonumber\\&&\hspace{1cm}\frac{q^{2}}{2|\vec{P_{_f}}|^2}(3\cos^{2}\theta-1)
+2\bigg[\frac{F_{_3}E_{_f}}{MM_{_f}}\bigg((C_{_{f_1}}+C^{'}_{_{f_1}})+\frac{P_{_f}q}{M^2_{_f}}(C_{_{f_3}}-C^{'}_{_{f_3}})\bigg)+\frac{F_{_1}}{M_{_f}}(C_{_{f_3}}+C^{'}_{_{f_3}})\bigg]
\nonumber\\&&\hspace{1cm}\frac{E_{_f}}{M}\frac{q^{4}}{8|\vec{P_{_f}}|^{2}}(5\cos^{4}\theta-6\cos^{2}\theta+1)
+2\bigg[\frac{F_{_1}}{M_{_f}}\bigg((C_{_{f_1}}-C^{'}_{_{f_1}})+\frac{P_{_f}q}{M^2_{_f}}(C_{_{f_3}}+C^{'}_{_{f_3}})\bigg)\bigg]\frac{q^{3}}{2|\vec{P_{_f}}|}\times
\nonumber\\&&\hspace{1cm}(\cos^{3}\theta-\cos\theta)-\bigg[\frac{F_{_3}E_{_f}}{MM_{_f}}\bigg((C_{_{f_1}}+C^{'}_{_{f_1}})+\frac{P_{_f}q}{M^2_{_f}}(C_{_{f_3}}-C^{'}_{_{f_3}})\bigg)
+\frac{F_{_1}}{M_{_f}}(C_{_{f_3}}+C^{'}_{_{f_3}})\bigg]\times
\nonumber\\&&\hspace{1cm}\frac{q^{4}}{8|\vec{P_{_f}}|^{2}}(5\cos^{4}\theta-6\cos^{2}\theta+1)\bigg\}.
\end{eqnarray}
\begin{eqnarray}
&&s_{_4}=\int\frac{q^2~d{q}~d{\cos\theta}}{(2\pi)^{2}}4\bigg\{\bigg[\frac{F_{_1}q^2}{M_{_f}}\bigg((C_{_{f_1}}-C^{'}_{_{f_1}})+
\frac{\alpha_{_{f}}|\vec{P_{_f}}|^2}{M^2_{_f}}(C_{_{f_3}}-C^{'}_{_{f_3}})\bigg)-\frac{F_{_3}\alpha^2_{_{f}}E_{_f}q^2}{MM_{_f}}\times
\nonumber\\&&\hspace{1cm}(C_{_{f_1}}-C^{'}_{_{f_1}})\bigg]\frac{q}{|\vec{P_{_f}}|}\cos\theta
+\bigg[\frac{F_{_1}}{M_{_f}}\bigg(-(C_{_{f_1}}-C^{'}_{_{f_1}})+\frac{\alpha_{_{f}}E^2_{_f}}{M^2_{_f}}(C_{_{f_3}}-C^{'}_{_{f_3}})\bigg)\bigg]\frac{q^{3}}{2|\vec{P_{_f}}|}\times
\nonumber\\&&\hspace{1cm}(\cos^{3}\theta-\cos\theta)+\bigg[\frac{F_{_3}E_{_f}}{MM_{_f}}\bigg(-(C_{_{f_1}}+C^{'}_{_{f_1}})+
\frac{P_{_f}q}{M^2_{_f}}(C_{_{f_3}}-C^{'}_{_{f_3}})\bigg)-\frac{F_{_1}}{M_{_f}}(C_{_{f_3}}+
\nonumber\\&&\hspace{1cm}C^{'}_{_{f_3}})\bigg]\frac{q^{4}}{8|\vec{P_{_f}}|^{2}}(5\cos^{4}\theta-6\cos^{2}\theta+1)+\bigg[-\frac{F_{_1}P_{_f}q}{M_{_f}}(C_{_{f_1}}-C^{'}_{_{f_1}})
+\frac{F_{_3}\alpha^2_{_{f}}E_{_f}P_{_f}q}{MM_{_f}}\times
\nonumber\\&&\hspace{1cm}(C_{_{f_1}}-C^{'}_{_{f_1}})\bigg]\frac{q^2}{2|\vec{P_{_f}}|^2}(3\cos^{2}\theta-1)\bigg\}.
\end{eqnarray}}

{\color{red}For the decay channel $X(3823)$ $\rightarrow$ $\chi_{_{c2}}(^3P_{_2})\gamma$, the form factors $g{_{_i}}$ are
\begin{eqnarray}
&&g_{_1}=\int\frac{q^2~d{q}~d{\cos\theta}}{(2\pi)^{2}}4\bigg\{\bigg[\frac{F_{_1}\alpha_{_{f}}E_{_f}}{MM_{_f}}(D_{_{f_7}}-D^{'}_{_{f_7}})+\frac{F_{_2}}{M}(D_{_{f_6}}-D^{'}_{_{f_6}})
-\frac{F_{_3}q^2}{M^2M_{_f}}(D_{_{f_7}}-D^{'}_{_{f_7}})\bigg]\times
\nonumber\\&&\hspace{1cm}\frac{E_{f}}{M}\frac{q^{3}}{2|\vec{P_{_f}}|^3}(5\cos^{3}\theta-3\cos\theta)+\bigg[-\frac{F_{_2}\alpha^2_{_{f}}E^2_{_f}}{M^3M^2_{_f}}(D_{_{f_4}}-D^{'}_{_{f_4}})+
\frac{F_{_3}\alpha^2_{_{f}}E^2_{_f}}{M^4M_{_f}}(D_{_{f_7}}-D^{'}_{_{f_7}})\bigg]\times
\nonumber\\&&\hspace{1cm}\frac{q^{3}}{2|\vec{P_{_f}}|}(\cos^{3}\theta-\cos\theta)+\bigg[-\frac{F_{_1}\alpha_{_{f}}E_{_f}M_{_f}}{M^3}(D_{_{f_5}}+
D^{'}_{_{f_5}})+\frac{F_{_3}\alpha^2_{_{f}}E^3_{_f}P_{_f}q}{M^5M_{_f}}(D_{_{f_7}}-D^{'}_{_{f_7}})\bigg)\bigg]\times
\nonumber\\&&\hspace{1cm}\frac{q^{2}}{2|\vec{P_{_f}}|^{2}}(3\cos^{2}\theta-1)+\bigg[\frac{F_{_1}M_{_f}}{M^2}(D_{_{f_5}}-D^{'}_{_{f_5}})-
\frac{F_{_3}\alpha_{_{f}}E_{_f}P_{_f}q}{M^3M_{_f}}(D_{_{f_7}}+D^{'}_{_{f_7}})\bigg]\frac{E_{_f}}{M}\frac{q^{3}}{2|\vec{P_{_f}}|^{3}}\times
\nonumber\\&&\hspace{1cm}(5\cos^{3}\theta-3\cos\theta)-\bigg[\frac{F_{_1}\alpha^2_{_{f}}E^2_{_f}}{M^3}(D_{_{f_7}}+D^{'}_{_{f_7}})-
\frac{F_{_2}\alpha_{_{f}}E_{_f}}{M^2}(D_{_{f_6}}+D^{'}_{_{f_6}})+\frac{F_{_3}\alpha_{_{f}}E_{_f}q^2}{M^3M_{_f}}\times
\nonumber\\&&\hspace{1cm}(D_{_{f_7}}+D^{'}_{_{f_7}})\bigg]\frac{q^{2}}{2|\vec{P_{_f}}|^2}(3\cos^{2}\theta-1)
-2\bigg[-\frac{F_{_1}\alpha_{_{f}}}{M^2M_{_f}}(D_{_{f_3}}+D^{'}_{_{f_3}})+\frac{F_{_2}\alpha_{_{f}}E_{_f}}{M^2M^2_{_f}}(D_{_{f_4}}+D^{'}_{_{f_4}})
\nonumber\\&&\hspace{1cm}-\frac{F_{_3}\alpha_{_{f}}E_{_f}}{M^3M_{_f}}(D_{_{f_7}}+D^{'}_{_{f_7}})\bigg]\frac{E_{_f}}{M}\frac{q^{4}}{8|\vec{P_{_f}}|^{2}}(5\cos^{4}\theta-6\cos^{2}\theta+1)-
\bigg[-\frac{2F_{_3}\alpha^2_{_{f}}E_{_f}}{M^3M_{_f}}(D_{_{f_3}}+D^{'}_{_{f_3}})
\nonumber\\&&\hspace{1cm}-\frac{F_{_1}}{M^2M_{_f}}(D_{_{f_7}}+D^{'}_{_{f_7}})+2\frac{F_{_2}\alpha_{_{f}}E_{_f}}{M^2M^2_{_f}}(D_{_{f_4}}+D^{'}_{_{f_4}})\bigg]
\frac{E_{_f}}{M}\frac{q^{4}}{8|\vec{P_{_f}}|^{2}}(5\cos^{4}\theta-6\cos^{2}\theta+1)
\nonumber\\&&\hspace{1cm}+2\bigg[\frac{F_{_3}\alpha_{_{f}}}{M^2M_{_f}}(D_{_{f_3}}-D^{'}_{_{f_3}})-\frac{F_{_2}}{MM^2_{_f}}(D_{_{f_4}}-D^{'}_{_{f_4}})\bigg]\frac{E^2_{_f}}{M^2}
\frac{q^{5}}{8|\vec{P_{_f}}|^{3}}(7\cos^{5}\theta-10\cos^{3}\theta+
\nonumber\\&&\hspace{1cm}3\cos\theta)-\bigg[\frac{F_{_3}M_{f}}{M^2}(D_{_{f_5}}+D^{'}_{_{f_5}})+\frac{F_{_1}E_{_f}}{MM_{_f}}(D_{_{f_7}}+D^{'}_{_{f_7}})\bigg]
\frac{1}{M^2}\frac{q^{4}}{8|\vec{P_{_f}}|^{2}}(5\cos^{4}\theta-6\cos^{2}\theta
\nonumber\\&&\hspace{1cm}+1)-2\bigg[\frac{F_{_3}\alpha_{_{f}}}{M^2M_{_f}}(D_{_{f_3}}-D^{'}_{_{f_3}})
-\frac{F_{_2}}{MM^2_{_f}}(D_{_{f_4}}-D^{'}_{_{f_4}})\bigg]\frac{1}{M^2}\frac{q^{5}}{8|\vec{P_{_f}}|}(\cos^{5}\theta-2\cos^{3}\theta+
\nonumber\\&&\hspace{1cm}\cos\theta)+\bigg[\frac{F_{_3}\alpha_{_{f}}E_{_f}M_{f}}{M^3}
(D_{_{f_5}}-D^{'}_{_{f_5}})+\frac{F_{_1}\alpha_{_{f}}E^2_{_f}}{M^2M_{_f}}(D_{_{f_7}}-D^{'}_{_{f_7}})\bigg](\frac{E_{_f}}{M}-1)\frac{q^{3}}{2|\vec{P_{_f}}|^{3}}(5\cos^{3}\theta
\nonumber\\&&\hspace{1cm}-3\cos\theta)+2\bigg[-\frac{F_{_1}\alpha^2_{_{f}}E_{_f}}{M^3M_{_f}}(D_{_{f_3}}+D^{'}_{_{f_3}})-\frac{F_{_3}}{M^2M_{_f}}(D_{_{f_7}}+D^{'}_{_{f_7}})+
\frac{F_{_2}\alpha_{_{f}}E_{_f}}{M^2M^2_{_f}}(D_{_{f_4}}+D^{'}_{_{f_4}})
\nonumber\\&&\hspace{1cm}\bigg]\frac{q^{4}}{8|\vec{P_{_f}}|^{2}}(5\cos^{4}\theta-6\cos^{2}\theta+1)-2\bigg[\frac{F_{_3}\alpha_{_{f}}}{M^2M_{_f}}(D_{_{f_3}}-D^{'}_{_{f_3}})
-\frac{F_{_2}}{MM^2_{_f}}(D_{_{f_4}}-D^{'}_{_{f_4}})\bigg]\times
\nonumber\\&&\hspace{1cm}\frac{E_{_f}}{M}\frac{q^{5}}{8|\vec{P_{_f}}|^{3}}(7\cos^{5}\theta-10\cos^{3}\theta+3\cos\theta)\bigg\}.
\end{eqnarray}
\begin{eqnarray}
&&g_{_2}=\int\frac{q^2~d{q}~d{\cos\theta}}{(2\pi)^{2}}4\bigg\{-2\bigg[\frac{F_{_3}\alpha_{_{f}}}{M^2M_{_f}}(D_{_{f_3}}-D^{'}_{_{f_3}})-\frac{F_{_2}}{MM^2_{_f}}(D_{_{f_4}}
-D^{'}_{_{f_4}})\bigg]\frac{E_{_f}}{M}\frac{q^{5}}{8|\vec{P_{_f}}|}(\cos^{5}\theta-
\nonumber\\&&\hspace{1cm}2\cos^{3}\theta+\cos\theta)+\bigg[-\frac{F_{_1}\alpha_{_{f}}}{M^2M_{_f}}(D_{_{f_3}}+D^{'}_{_{f_3}})+\frac{F_{_2}\alpha_{_{f}}E_{_f}}{M^2M^2_{_f}}(D_{_{f_4}}+D^{'}_{_{f_4}})
-\frac{F_{_3}\alpha_{_{f}}E_{_f}}{M^3M_{_f}}(D_{_{f_7}}+
\nonumber\\&&\hspace{1cm}D^{'}_{_{f_7}})\bigg]\frac{q^{4}}{8}(\cos^{4}\theta-2\cos^{2}\theta+1)
-\bigg[\frac{F_{_1}M_{_f}}{M^2}(D_{_{f_5}}-D^{'}_{_{f_5}})-\frac{F_{_3}\alpha_{_{f}}E_{_f}P_{_f}q}{M^3M_{_f}}(D_{_{f_7}}+D^{'}_{_{f_7}})\bigg]\times
\nonumber\\&&\hspace{1cm}\frac{q^{3}}{2|\vec{P_{_f}}|}(\cos^{3}\theta-\cos\theta)-\bigg[\frac{1}{MM_{_f}}(\alpha_{_{f}}F_{_1}E_{_f}-\frac{F_{_3}q^2}{M})(D_{_{f_7}}-D^{'}_{_{f_7}})+
\frac{F_{_2}}{M}(D_{_{f_6}}-D^{'}_{_{f_6}})\bigg]\times
\nonumber\\&&\hspace{1cm}\frac{E_{_f}}{M}\frac{q^{3}}{2|\vec{P_{_f}}|}(\cos^{3}\theta-\cos\theta)-\bigg[\frac{F_{_3}M_{f}}{M^2}(D_{_{f_5}}+D^{'}_{_{f_5}})+
\frac{F_{_1}E_{_f}}{MM_{_f}}(D_{_{f_7}}+D^{'}_{_{f_7}})\bigg]\frac{E_{_f}}{M}\frac{q^{4}}{8|\vec{P_{_f}}|^{2}}\times
\nonumber\\&&\hspace{1cm}(5\cos^{4}\theta-6\cos^{2}\theta+1)+\bigg[\frac{F_{_3}M_{f}}{M^2}(D_{_{f_5}}+D^{'}_{_{f_5}})+
\frac{F_{_1}E_{_f}}{MM_{_f}}(D_{_{f_7}}+D^{'}_{_{f_7}})\bigg]\frac{q^{4}}{8|\vec{P_{_f}}|^{2}}\times
\nonumber\\&&\hspace{1cm}(5\cos^{4}\theta-6\cos^{2}\theta+1)\bigg\}.
\end{eqnarray}
\begin{eqnarray}
&&g_{_4}=\int\frac{q^2~d{q}~d{\cos\theta}}{(2\pi)^{2}}4\bigg\{\bigg[\frac{F_{_3}M_{f}}{M^2}(D_{_{f_5}}+D^{'}_{_{f_5}})+\frac{F_{_1}}{M}\frac{E_{_f}}{M_{_f}}(D_{_{f_7}}+D^{'}_{_{f_7}})\bigg]
\frac{q^{4}}{8|\vec{P_{_f}}|^{2}}(5\cos^{4}\theta-
\nonumber\\&&\hspace{1cm}6\cos^{2}\theta+1)+2\bigg[\frac{F_{_3}\alpha_{_{f}}}{M^2M_{_f}}(D_{_{f_3}}-D^{'}_{_{f_3}})-\frac{F_{_2}}{MM^2_{_f}}(D_{_{f_4}}-D^{'}_{_{f_4}})\bigg]\frac{E_{_f}}{M}
\frac{q^{5}}{8|\vec{P_{_f}}|}(\cos^{5}\theta-
\nonumber\\&&\hspace{1cm}2\cos^{3}\theta+\cos\theta)-\bigg[(\frac{F_{_1}\alpha_{_{f}}E_{_f}}{MM_{_f}}+\frac{F_{_3}q^2}{M^2M_{_f}})(D_{_{f_7}}-D^{'}_{_{f_7}})+\frac{F_{_2}}{M}
(D_{_{f_4}}-D^{'}_{_{f_4}})\bigg]\frac{q^{3}}{2|\vec{P_{_f}}|}\times
\nonumber\\&&\hspace{1cm}(\cos^{3}\theta-\cos\theta)\bigg\}.
\end{eqnarray}
\begin{eqnarray}
&&g_{_5}=\int\frac{q^2~d{q}~d{\cos\theta}}{(2\pi)^{2}}4\bigg\{\bigg[\frac{F_{_1}\alpha_{_{f}}E_{_f}M_{_f}}{M^3}(D_{_{f_5}}+D^{'}_{_{f_5}})
-\frac{F_{_2}\alpha_{_{f}}E^2_{_f}}{M^3}(D_{_{f_1}}+D^{'}_{_{f_1}})+\frac{F_{_3}\alpha^2_{_{f}}E^2_{_f}}{M^4M_{_f}}\times
\nonumber\\&&\hspace{1cm}\bigg(q^2(D_{_{f_3}}+D^{'}_{_{f_3}})+\alpha_{_{f}}P_{_f}q(D_{_{f_3}}-D^{'}_{_{f_3}})\bigg)\bigg]\frac{q^{2}}{2|\vec{P_{_f}}|^{2}}(3\cos^{2}\theta-1)
+\bigg[\frac{F_{_3}\alpha^3_{_{f}}E^2_{_f}}{M^4M_{_f}}(D_{_{f_3}}-
\nonumber\\&&\hspace{1cm}D^{'}_{_{f_3}})+\frac{F_{_1}\alpha_{_{f}}E_{_f}}{M^3M_{_f}}(D_{_{f_7}}-D^{'}_{_{f_7}})-\frac{F_{_2}\alpha^2_{_{f}}E^2_{_f}}{M^3M^2_{_f}}(D_{_{f_4}}-D^{'}_{_{f_4}})\bigg]
\frac{q^{3}}{2|\vec{P_{_f}}|}(\cos^{3}\theta-\cos\theta)+
\nonumber\\&&\hspace{1cm}\bigg[-\frac{F_{_2}}{M}(D_{_{f_1}}+D^{'}_{_{f_1}})+\frac{F_{_3}}{M^2}\bigg(\frac{q^2}{M_{_f}}(D_{_{f_3}}+D^{'}_{_{f_3}})+\frac{P_{_f}q}{M_{_f}}(D_{_{f_3}}-D^{'}_{_{f_3}})
+M_{_f}(D_{_{f_5}}+D^{'}_{_{f_5}})
\nonumber\\&&\hspace{1cm}\bigg)\bigg]\frac{1}{M^2}\frac{q^{4}}{8|\vec{P_{_f}}|^{2}}(5\cos^{4}\theta-6\cos^{2}\theta+1)+\bigg[\frac{F_{_3}\alpha_{_{f}}}{M^2M_{_f}}
(D_{_{f_3}}-D^{'}_{_{f_3}})-\frac{F_{_2}}{MM^2_{_f}}(D_{_{f_4}}-D^{'}_{_{f_4}})\bigg]
\nonumber\\&&\hspace{1cm}\frac{E^2_{_f}}{M^2}\frac{q^{5}}{8|\vec{P_{_f}}|^3}(7\cos^{5}\theta-10\cos^{3}\theta+3\cos\theta)
+\bigg[\frac{F_{_2}\alpha_{_{f}}E_{_f}}{M^2}(D_{_{f_1}}-D^{'}_{_{f_1}})-\frac{F_{_3}\alpha_{_{f}}E_{_f}}{M^3}\times
\nonumber\\&&\hspace{1cm}\bigg(\frac{q^2}{M_{_f}}(D_{_{f_3}}-D^{'}_{_{f_3}})+\frac{\alpha_{_{f}}P_{_f}q}{M_{_f}}(D_{_{f_3}}+D^{'}_{_{f_3}})
+\alpha_{_{f}}M_{_f}(D_{_{f_5}}-D^{'}_{_{f_5}})+\alpha_{_{f}}M_{_f}(D_{_{f_7}}-D^{'}_{_{f_7}})\bigg)
\nonumber\\&&\hspace{1cm}-\frac{F_{_1}}{M^2}\bigg(M_{_f}(D_{_{f_5}}-D^{'}_{_{f_5}})+\frac{P_{_f}q}{M_{_f}}(D_{_{f_7}}+D^{'}_{_{f_7}})\bigg)\bigg]
\frac{E_{_f}}{M}\frac{q^{3}}{2|\vec{P_{_f}}|^{3}}(5\cos^{3}\theta-3\cos\theta)-\bigg[F_{_3}\times
\nonumber\\&&\hspace{1cm}\frac{\alpha_{_{f}}}{M^2M_{_f}}\times(D_{_{f_3}}-D^{'}_{_{f_3}})-\frac{F_{_2}}{MM^2_{_f}}(D_{_{f_4}}-D^{'}_{_{f_4}})\bigg]
\frac{1}{M^2}\frac{q^{5}}{8|\vec{P_{_f}}|}(\cos^{5}\theta-2\cos^{3}\theta+\cos\theta)+
\nonumber\\&&\hspace{1cm}\bigg[2\frac{F_{_3}\alpha^2_{_{f}}E_{_f}}{M^3M_{_f}}(D_{_{f_3}}+D^{'}_{_{f_3}})+\frac{F_{_1}}{M^2M_{_f}}(D_{_{f_7}}+D^{'}_{_{f_7}})
-\frac{2F_{_2}\alpha_{_{f}}E_{_f}}{M^2M^2_{_f}}(D_{_{f_4}}+D^{'}_{_{f_4}})\bigg]\frac{E_{_f}}{M}\frac{q^{4}}{8|\vec{P_{_f}}|^{2}}\times
\nonumber\\&&\hspace{1cm}(5\cos^{4}\theta-6\cos^{2}\theta+1)\bigg\}.
\end{eqnarray}
\begin{eqnarray}
&&g_{_6}=\int\frac{q^2~d{q}~d{\cos\theta}}{(2\pi)^{2}}4\bigg\{\bigg[-\frac{F_{_1}\alpha^2_{_{f}}E_{_f}}{M^3M_{_f}}(D_{_{f_3}}+D^{'}_{_{f_3}})-\frac{F_{_3}}{M^2M_{_f}}(D_{_{f_7}}+D^{'}_{_{f_7}})
+\frac{F_{_2}\alpha_{_{f}}E_{_f}}{M^2M^2_{_f}}\times
\nonumber\\&&\hspace{1cm}(D_{_{f_4}}+D^{'}_{_{f_4}})\bigg]\frac{q^{4}}{8}(\cos^{4}\theta-2\cos^{2}\theta+1)-
\bigg[\frac{F_{_3}\alpha_{_{f}}}{M^2M_{_f}}(D_{_{f_3}}-D^{'}_{_{f_3}})-\frac{F_{_2}}{MM^2_{_f}}(D_{_{f_4}}-
\nonumber\\&&\hspace{1cm}D^{'}_{_{f_4}})\bigg]\frac{E_{_f}}{M}\frac{q^{5}}{8|\vec{P_{_f}}|}(\cos^{5}\theta-2\cos^{3}\theta+\cos\theta)
-\bigg[-\frac{F_{_1}}{M^2}\bigg(M_{_f}(D_{_{f_5}}-D^{'}_{_{f_5}})+\frac{P_{_f}q}{M_{_f}}\times
\nonumber\\&&\hspace{1cm}(D_{_{f_7}}+D^{'}_{_{f_7}})\bigg)+\frac{F_{_2}\alpha_{_{f}}E_{_f}}{M^2}(D_{_{f_1}}-D^{'}_{_{f_1}})-
\frac{F_{_3}\alpha_{_{f}}E_{_f}}{M^3}\bigg(\frac{q^2}{M_{_f}}(D_{_{f_3}}-D^{'}_{_{f_3}})+\frac{\alpha_{_{f}}P_{_f}q}{M_{_f}}\times
\nonumber\\&&\hspace{1cm}(D_{_{f_3}}+D^{'}_{_{f_3}})+\alpha_{_{f}}M_{_f}(D_{_{f_5}}-D^{'}_{_{f_5}})+\alpha_{_{f}}M_{_f}(D_{_{f_7}}-D^{'}_{_{f_7}})\bigg)\bigg]
\frac{q^{3}}{2|\vec{P_{_f}}|}(\cos^{3}\theta-\cos\theta)\bigg\}.
\nonumber\\&&\hspace{1cm}
\end{eqnarray}
\begin{eqnarray}
&&g_{_7}=\int\frac{q^2~d{q}~d{\cos\theta}}{(2\pi)^{2}}4\bigg\{-\bigg[\frac{F_{_3}\alpha_{_{f}}E_{_f}M_{f}}{M^3}(D_{_{f_5}}-D^{'}_{_{f_5}})
+\frac{F_{_1}\alpha_{_{f}}E^2_{_f}}{M^2M_{_f}}(D_{_{f_7}}-D^{'}_{_{f_7}})\bigg]\frac{q^{3}}{2|\vec{P_{_f}}|}\times
\nonumber\\&&\hspace{1cm}(\cos^{3}\theta-\cos\theta)+2\bigg[-\frac{F_{_2}}{M}(D_{_{f_1}}+D^{'}_{_{f_1}})+
\frac{F_{_3}}{M^2}\bigg(\frac{q^2}{M_{_f}}(D_{_{f_3}}+D^{'}_{_{f_3}})+\frac{P_{_f}q}{M_{_f}}\times
\nonumber\\&&\hspace{1cm}(D_{_{f_3}}-D^{'}_{_{f_3}})+M_{_f}(D_{_{f_5}}+D^{'}_{_{f_5}})\bigg)\bigg]\frac{q^{4}}{8|\vec{P_{_f}}|^{2}}(5\cos^{4}\theta-6\cos^{2}\theta+1)-\bigg[-\frac{F_{_1}}{M^2}\times
\nonumber\\&&\hspace{1cm}\bigg(M_{_f}(D_{_{f_5}}-D^{'}_{_{f_5}})+\frac{P_{_f}q}{M_{_f}}(D_{_{f_7}}+D^{'}_{_{f_7}})\bigg)+\frac{F_{_2}\alpha_{_{f}}E_{_f}}{M^2}(D_{_{f_1}}-D^{'}_{_{f_1}})
-2\frac{F_{_3}\alpha_{_{f}}E_{_f}}{M^3}\times
\nonumber\\&&\hspace{1cm}\bigg(\frac{q^2}{M_{_f}}(D_{_{f_3}}-D^{'}_{_{f_3}})+\frac{\alpha_{_{f}}P_{_f}q}{M_{_f}}(D_{_{f_3}}+D^{'}_{_{f_3}})+
\alpha_{_{f}}M_{_f}(D_{_{f_5}}-D^{'}_{_{f_5}})+\alpha_{_{f}}M_{_f}(D_{_{f_7}}-
\nonumber\\&&\hspace{1cm}D^{'}_{_{f_7}})\bigg)\bigg]\frac{q^{3}}{2|\vec{P_{_f}}|}(\cos^{3}\theta-\cos\theta)-\bigg[\frac{F_{_3}M_{f}}{M^2}
(D_{_{f_5}}+D^{'}_{_{f_5}})+\frac{F_{_1}E_{_f}}{MM_{_f}}(D_{_{f_7}}+D^{'}_{_{f_7}})\bigg]\times
\nonumber\\&&\hspace{1cm}\frac{E_{_f}}{M}\frac{q^{4}}{8|\vec{P_{_f}}|^{2}}\times(5\cos^{4}\theta-6\cos^{2}\theta+1)\bigg\}.
\end{eqnarray}}

\section{The amplitude square }
For $X(3823)$ $\rightarrow$ $\chi_{_{c1}}(^3P_{_1})\gamma$, the square modulus of amplitude is
\begin{eqnarray}
&&\overline{{|\cal{M}|}^2_3}=\frac{4e^2}{45}\bigg\{2s^2_{_1}M^2\bigg[(|\vec{P_{_f}}|^{4}+\frac{M^6_{_f}-E^6_{_f}}{3M^2_{_f}})(2ME_{_f}+M^2+M^2_{_f})\bigg]+\frac{5}{3}s^2_{_2}M^2\frac{|\vec{P_{_f}}|^{4}}{M^2_{_f}}
\nonumber\\&&\hspace{1.2cm}+s^2_{_3}\bigg[M^2_{_f}(2ME_{_f}+M^2+M^2_{_f})-\frac{1}{3}\bigg(E^2_{_f}(E^2_{_f}+M^2+|\vec{P_{_f}}|^{2})-\frac{2ME^3_{_f}|\vec{P_{_f}}|^{2}}{M^2_{_f}}
\nonumber\\&&\hspace{1.2cm}+\frac{2ME^4_{_f}(E_{_f}+M)}{M^2_{f}}\bigg)\bigg]+2s^2_{_4}|\vec{P_{_f}}|^{4}+\frac{5}{3}s^2_{_5}(\frac{E^2_{_f}}{M^2_{_f}}+2)+4s_{_1}s_{_2}M^2\bigg(E^2_{_f}|\vec{P_{_f}}|^{2}+
\nonumber\\&&\hspace{1.2cm}\frac{(M^6_{_f}-E^6_{_f})}{3M^2_{_f}}\bigg)+\frac{4}{3}s_{_1}s_{_3}ME_{_f}\bigg(\frac{|\vec{P_{_f}}|^{4}}{M^2_{_f}}(2ME_{_f}+M^2+M^2_{_f})\bigg)+4s_{_1}s_{_4}M^2\times
\nonumber\\&&\hspace{1.2cm}\bigg(E^2_{_f}|\vec{P_{_f}}|^{2}+\frac{(M^6_{_f}-E^6_{_f})}{3M^2_{_f}}\bigg)+(s_{_1}s_{_5}+s_{_2}s_{_3}+s_{_2}s_{_4}+s_{_3}s_{_4})\frac{4}{3}\frac{ME_{_f}|\vec{P_{_f}}|^{4}}{M^2_{_f}}
\nonumber\\&&\hspace{1.2cm}+\frac{10}{3}s_{_2}s_{_5}\frac{ME_{_f}|\vec{P_{_f}}|^{2}}{M^2_{_f}}+\frac{2}{3}s_{_3}s_{_5}|\vec{P_{_f}}|^{2}\bigg(1+\frac{2(M^2_{_f}+E^2_{_f})}{M^2_{_f}}\bigg)+\frac{4}{3}s_{_4}s_{_5}\frac{|\vec{P_{_f}}|^{4}}{M^2_{_f}}\bigg\}.
\end{eqnarray}

For $X(3823)$ $\rightarrow$ $\chi_{_{c2}}(^3P_{_2})\gamma$, the square of the amplitude is
\begin{eqnarray}
&&\overline{{|\cal{M}|}^2_4}=\frac{4e^2}{45}\bigg\{g^2_{_1}M^4\bigg[\frac{3}{2}|\vec{P_{_f}}|^{2}E^2_{_f}(2ME_{_f}+M^2+M^2_{_f})+\frac{ME_{_f}(M^6_{_f}-E^6_{_f})}{M^2_{_f}}\bigg]+\frac{2}{3}g^2_{_5}M^6\times
\nonumber\\&&\hspace{1.5cm}(|\vec{P_{_f}}|^{6}\frac{E^2_{_f}}{M^4_{_f}}+\frac{M^6_{_f}-E^6_{_f}}{M^2_{_f}})+\frac{1}{2}g^2_{_3}M^2\bigg(3|\vec{P_{_f}}|^{2}(2ME_{_f}+M^2+M^2_{_f})+(M^4_{_f}-E^4_{_f})\times
\nonumber\\&&\hspace{1.5cm}\frac{2ME_{_f}+M^2}{M^2_{_f}}\bigg)+\frac{5}{3}g^2_{_7}M^2|\vec{P_{_f}}|^{4}+\frac{1}{6}g^2_{_8}\frac{M^2|\vec{P_{_f}}|^{2}}{M^2_{_f}}(19+\frac{3E^2_{_f}}{M^2_{_f}})+\frac{1}{6}(g^2_{_2}+g^2_{_6})M^4|\vec{P_{_f}}|^{2}
\nonumber\\&&\hspace{1.5cm}\times\bigg(\frac{3E^2_{_f}(M^2_{_f}+E^2_{_f})}{M^4_{_f}}+\frac{E^2_{_f}}{M^2_{_f}}-7\bigg)+g_{_1}g_{_3}M^3E_{_f}|\vec{P_{_f}}|^{4}(1+\frac{M^2}{M^2_{_f}}+\frac{2ME_{_f}}{M^2_{_f}})
\nonumber\\&&\hspace{1.5cm}+(g_{_1}g_{_2}+g_{_1}g_{_4}+g_{_1}g_{_7})M^4(3E^2_{_f}|\vec{P_{_f}}|^{2}+\frac{M^6_{_f}-E^6_{_f}}{M^2_{_f}})+\frac{4}{3}(g_{_2}g_{_5}-g_{_5}g_{_6})\times
\nonumber\\&&\hspace{1.5cm}\frac{-M^5E_{_f}|\vec{P_{_f}}|^{6}}{M^4_{_f}}+\frac{2}{3}g_{_5}g_{_7}M^4(-3E^2_{_f}|\vec{P_{_f}}|^{2}+\frac{E^6_{_f}-M^6_{_f}}{M^2_{_f}})+\frac{2}{3}(g_{_5}g_{_8}+g_{_2}g_{_6})\frac{M^4|\vec{P_{_f}}|^{2}}{M^2_{_f}}
\nonumber\\&&\hspace{1.5cm}\times(M^2_{_f}+E^2_{_f}-\frac{2E^4_{_f}}{M^2_{_f}})+\frac{1}{3}(g_{_2}g_{_8}-g_{_6}g_{_8})\frac{M^3E_{_f}|\vec{P_{_f}}|^{2}}{M^2_{_f}}\bigg(\frac{4(M^2_{f}+E^2_{_f})}{M^2_{_f}}+7\bigg)
\nonumber\\&&\hspace{1.5cm}+(g_{_3}g_{_4}+g_{_2}g_{_3}+g_{_3}g_{_7}+g_{_1}g_{_8}+g_{_4}g_{_7}+g_{_2}g_{_7}+g_{_6}g_{_7})M^3E_{_f}\frac{-|\vec{P_{_f}}|^{4}}{M^2_{_f}}+g_{_3}g_{_8}M^2\times
\nonumber\\&&\hspace{1.5cm}(-3|\vec{P_{_f}}|^{2}+\frac{-|\vec{P_{_f}}|^{2}(M^2_{_f}+E^2_{_f})}{M^2_{_f}})+\frac{5}{3}(g_{_2}g_{_4}+g_{_4}g_{_6}+g^2_{_4}+\frac{g_{_7}g_{_8}}{M^2})M^4\frac{|\vec{P_{_f}}|^{4}}{M^2_{_f}}\bigg\}.
\end{eqnarray}

\end{document}